\documentclass[aps,nofootinbib,superscriptaddress,showpacs,amssymb,amsmath,amsfonts,altaffilletter,10pt,twocolumn,floatfix]{revtex4}
\usepackage{dcolumn} 
\usepackage{color}
\usepackage{url}
\usepackage{hyperref}
\usepackage{breakurl}
\usepackage{graphicx}
\def\gsim{\mathrel{
\rlap{\raise 0.5ex \hbox{$>$}}{\lower 0.6ex
\hbox{$\sim$}}}}
\def\lsim{\mathrel{
\rlap{\raise 0.5ex \hbox{$<$}}{\lower 0.6ex
\hbox{$\sim$}}}}

\begin{document}
\def\Cardiff{School of Physics and Astronomy, Cardiff University, 5 The Parade, Cardiff, CF24 3AA, UK} 
\def\Amsterdam{Nikhef -- National Institute for Subatomic Physics,
Science Park 105, 1098 XG Amsterdam, The Netherlands}
\def\Nice{UMR ARTEMIS, CNRS, University of Nice Sophia-Antipolis, Observatoire de la C\^ote d'Azur, BP
4229, 06304, Nice Cedex 4, France}
\def\Milwaukee{University of Wisconsin-Milwaukee P.O.\ Box 413, 2200 E.\ Kenwood Blvd.\  
Milwaukee, WI 53201-0413, USA}
\def\Maryland{Maryland Center for Fundamental Physics, Department of Physics, \\
University of Maryland, College Park, MD 20742, USA}
\title{A Mock Data Challenge for the Einstein Gravitational-Wave Telescope}
\author{Tania Regimbau}
\email{Tania.Regimbau@oca.eu\\ http://www.oca.eu/regimbau/ET-MDC_web/ET-MDC.html}
\affiliation{\Nice}
\author{Thomas Dent} 
\email{Thomas.Dent@astro.cf.ac.uk}
\affiliation{\Cardiff}
\author{Walter Del Pozzo}
\affiliation{\Amsterdam}
\author{Stefanos Giampanis}
\affiliation{\Milwaukee}
\author{Tjonnie G\,.F.\ Li}
\affiliation{\Amsterdam}
\author{Craig Robinson}
\affiliation{\Cardiff}
\affiliation{\Maryland}
\author{Chris Van Den Broeck}
\affiliation{\Amsterdam}
\author{Duncan Meacher}
\affiliation{\Cardiff}
\author{Carl Rodriguez}
\affiliation{\Cardiff}
\author{B.\,S.\ Sathyaprakash}
\affiliation{\Cardiff}
\author{Katarzyna W\'ojcik}
\affiliation{\Cardiff}

\begin{abstract}
Einstein Telescope (ET) is conceived to be a third generation gravitational-wave observatory.
Its amplitude sensitivity would be a factor ten better than advanced LIGO and Virgo and it 
could also extend the low-frequency sensitivity down to 1--3\,Hz, compared to the 10--20\,Hz 
of advanced detectors. Such an observatory will have the potential to observe a variety of 
different GW sources, including compact binary systems at cosmological distances. 
ET's expected reach for binary neutron star (BNS) coalescences is out to redshift $z\simeq 
2$ and the rate of detectable BNS coalescences could be as high as one every 
few tens or hundreds of seconds, each lasting up to several days. 
With such a signal-rich environment, a key question in data analysis is whether overlapping 
signals can be discriminated. 
In this paper we simulate the GW signals from a cosmological population of BNS and ask the
following questions: Does this population create a confusion background that limits ET's 
ability to detect foreground sources? How efficient are current algorithms in discriminating 
overlapping BNS signals? Is it possible to discern the presence of a population of
signals in the data by cross-correlating data from different detectors in the ET observatory? 
We find that algorithms currently used to analyze LIGO and Virgo data are already powerful 
enough to detect the sources expected in ET, but new algorithms are required to fully exploit 
ET data.
\end{abstract}
\maketitle

\section{Introduction}

After a decade of detector installation and commissioning, 
ground-based detectors looking for gravitational waves 
(GWs) have reached or surpassed their design sensitivities 
and are poised to open up a new window onto the Universe, 
as well as allowing coincident searches with electromagnetic 
or neutrino detectors.  The first generation of interferometric 
observatories (GEO \cite{GEO}, LIGO \cite{LIGO} and Virgo 
\cite{Virgo}) have already put interesting constraints, for 
example, on the ellipticity of the Crab pulsar \cite{crab} and 
on the cosmological stochastic background 
\cite{stochNature}. With the second generation starting in 
a few years, one expects to detect compact binary coalescences 
in the local Universe \cite{ratespaper}, while third 
generation detectors, such as the Einstein Telescope \cite{ET},
should take GW astronomy to a new level, due to the large 
numbers of high SNRs of detectable sources, making it possible 
to address a range of problems on a wide variety of 
astrophysical sources but also in fundamental physics and 
cosmology.

The coalescence of two neutron stars (BNS), two black holes 
(BBH) or a neutron star and a black hole (NS-BH), are the 
most promising sources for terrestrial detectors, due 
to the huge amount of energy emitted in the last phase of 
their inspiral trajectory, merger, and ringdown. The maximum
distance probed with current detectors is about 30\,Mpc 
\cite{Abadie:2010yb}
for BNS, but the next generation of detectors should be 
taking data with a sensitivity approximately 10 times 
greater, pushing the horizon up to about 450\,Mpc \cite{ratespaper}. 
With the 
third generation Einstein Telescope, the sensitivity will 
be increased by another order of magnitude and the horizon 
of compact binaries is expected to reach cosmological distances 
\cite{Sathyaprakash:2011bh}. Among other things, this will allow for a 
detailed study of the evolution of binary coalescences over 
redshift \cite{vdb10}, measurement of the mass function of
neutron stars and black holes and of the neutron star equation of 
state \cite{rmsucf09,Hinderer:2009ca}, and the use of binary 
neutron stars and neutron star-black hole binaries as 
standard sirens to constrain dark energy and its time 
evolution \cite{Sathyaprakash:2009xt,zvbl10,Messenger:2011gi} (for a summary of ET science
objectives see Ref.~\cite{Sathyaprakash:2011bh}). In such a large 
volume, however, the number of sources can be as large as a million 
and the waveforms may overlap to create a confusion foreground, 
especially at low frequencies where the signal can last 
for several days \cite{reg09}. This could affect our ability 
to make individual detections and perform parameter estimation, 
and the issue deserves thorough study. 

With this in mind, we have simulated Einstein Telescope detector 
noise and added signals from a population of compact binaries, 
with a view to issuing a Mock Data Challenge (MDC) to the 
gravitational-wave community. This could be used to develop 
advanced data analysis methods in order to separate the
sources and measure the properties of both individual sources 
and of the catalog as a whole. Initially we used a simple BNS 
inspiral signal model, but work is in progress to include other 
types of sources and more sophisticated scenarios. 
In the longer term we envisage issuing \emph{ET science 
challenges} to encompass not only detection of signals and 
parameter estimation, but also the application of such results to 
outstanding problems in fundamental physics, astrophysics and
cosmology.

In Section 2, we present the Einstein Telescope; in Section 3 
we describe our procedure to simulate the mock data; in Section 
4 we present the results of the search for both individual sources 
and the integrated signal; in Section 5 we discuss future developments
for the mock data and in the search methods; finally in Section 6 we 
draw our conclusions.
 
\section{Einstein Telescope}


\begin{figure*}
\centering
\includegraphics[angle=00,width=0.40\textwidth]{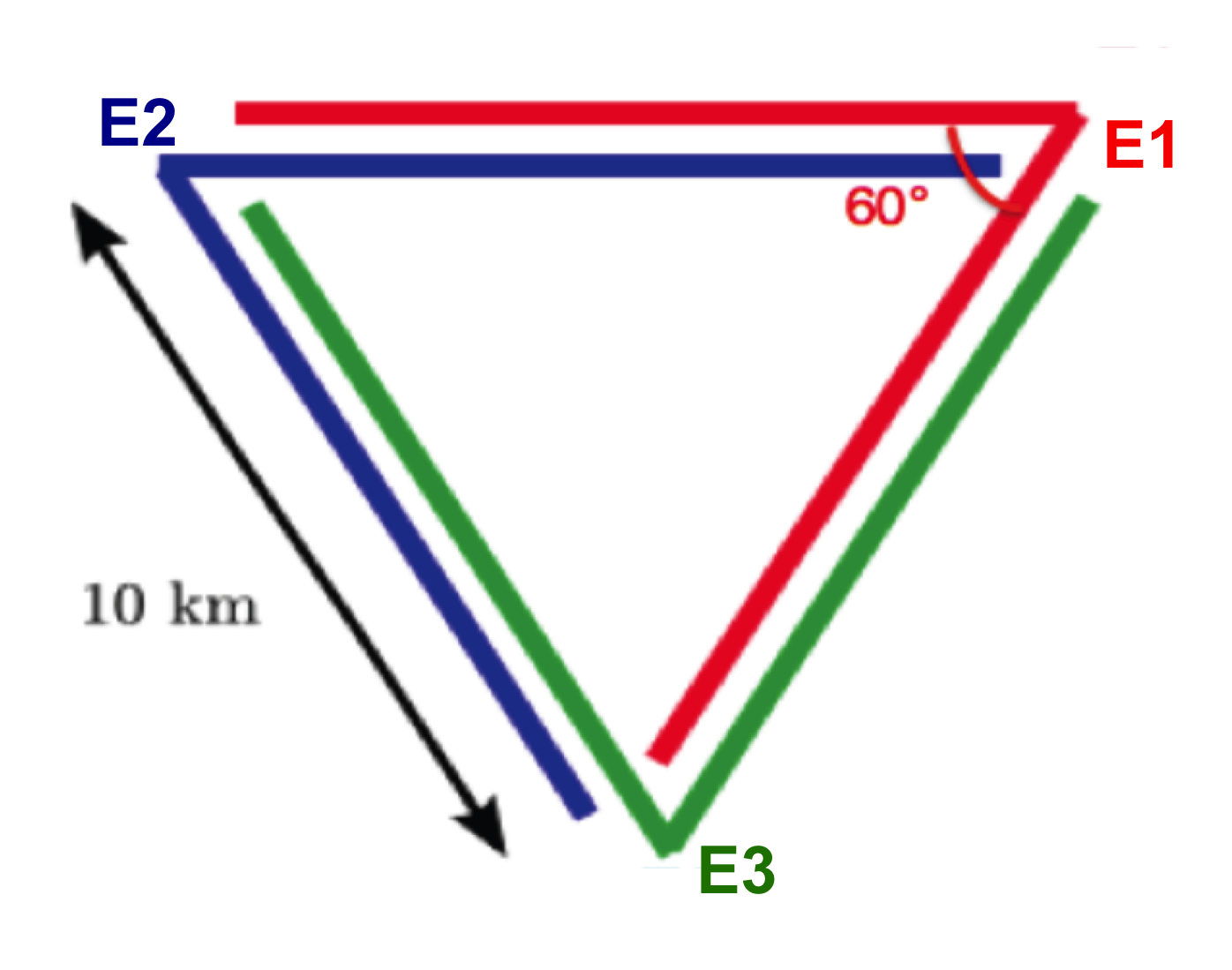}
\hskip0.1\textwidth
\includegraphics[angle=00,width=0.48\textwidth]{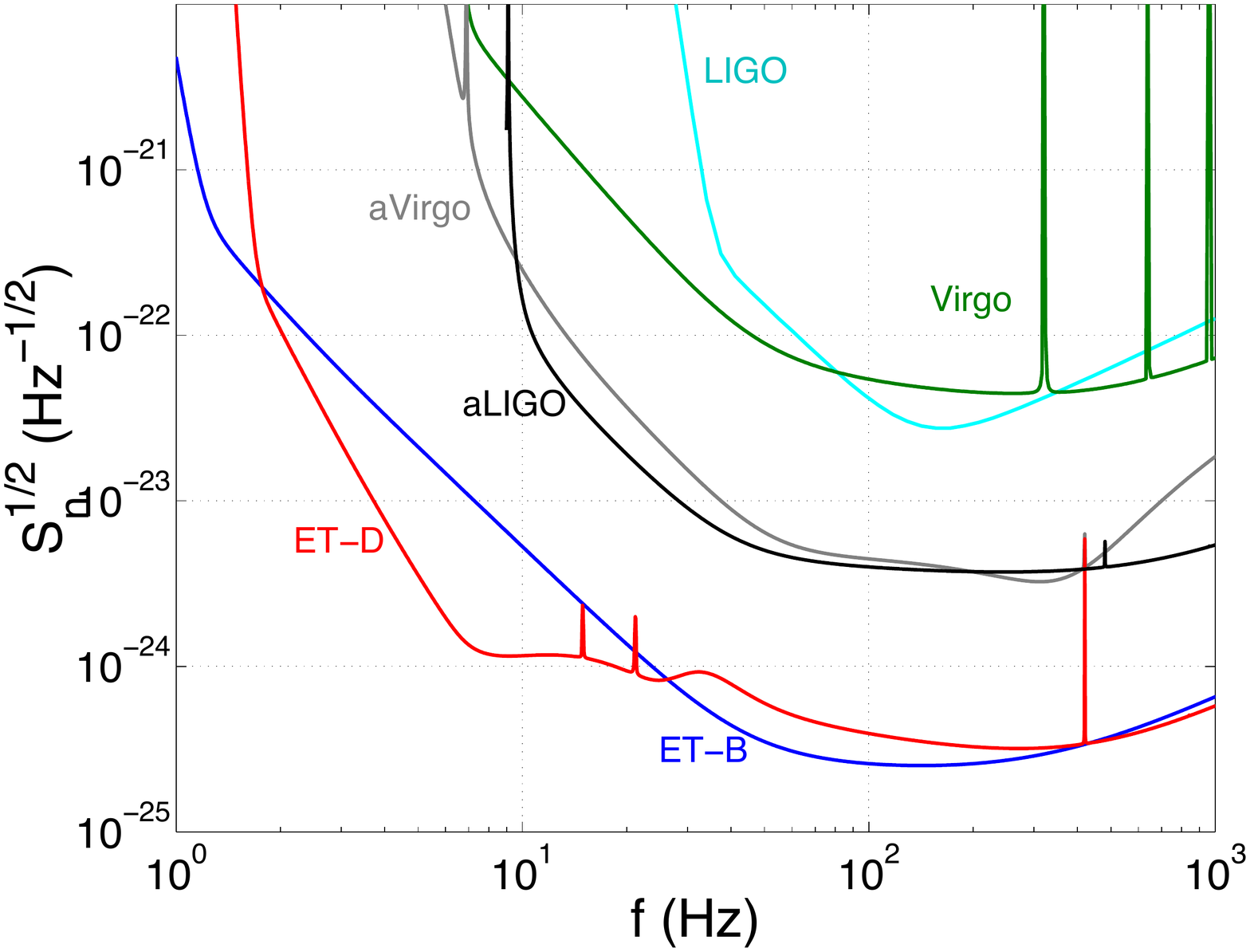}
\caption{Left : Schematic configuration of the planned GW detector 
Einstein Telescope. 
Right : sensitivity for the initial configuration, ET-B, considered in 
the Design Study \cite{DSD}, and the most evolved configuration ET-D, 
compared to the sensitivity of first generation detectors LIGO and Virgo 
and the projected sensitivity of second generation (advanced) 
detectors, here the aLIGO high-power zero detuning sensitivity 
\cite{aLIGO} and aVirgo \cite{aVirgo}.}
\label{fig-noise}
\label{fig-ET}
\end{figure*}

A conceptual design study was recently concluded for the proposed 
European project, the \emph{Einstein gravitational wave Telescope} 
(ET)~\cite{DSD}. The
goal of the study was to explore the technological challenges and
the scientific benefits of building a third generation gravitational
wave detector that is a factor 10 better than advanced detectors but
also capable of observing at frequencies down to 1--3\,Hz \cite{ET}. 
In this Section we will discuss the optical configuration of ET,
different design choices for its sensitivity and ET's response to
gravitational waves and its distance reach to compact binary 
coalescences. 

\subsection{Optical topology and sensitivity}
Consideration of many factors including continuous observation (duty 
cycle), ability to resolve the two polarizations of GW, and capacity to 
support new designs over many decades, led to the conclusion that the 
infrastructures housing the current detectors will be inadequate to meet 
the design goals of ET. The Einstein Telescope is envisioned to consist 
of three V-shaped Michelson interferometers with 60 degree opening 
angles, arranged in a triangle configuration (see Fig.~\ref{fig-ET}, left panel). 
These are to be placed underground to reduce the influence of seismic 
noise. 

The design goal to push the sensitivity floor at low frequency down
to 1--3\,Hz comes from the requirement that ET be sensitive to 
more massive coalescing binaries than advanced detectors, 
\emph{i.e.}\ intermediate mass BBH of masses in the range 
$10^2$--$10^4\,M_\odot$ \cite{Huerta:2010un,Huerta:2010tp,
Gair:2010dx,AmaroSeoane:2009ui}, but also be able to observe 
stellar mass binaries for far longer periods before they merge. 
With better low-frequency sensitivity, the subtle secular 
general-relativistic effects that occur in strong gravitational 
fields will accumulate over longer periods, as shown in Eq.~\eqref{eq-tau}, 
facilitating a deeper understanding of GW sources. Additionally, in the 
case of binaries where one or both components is a neutron star, the 
improved low-frequency sensitivity will allow the source's redshift to be 
measured \cite{Messenger:2011gi}, by breaking the degeneracy between
the redshifted mass measured from the PN phase and the rest-frame
mass measured from the NS tidal deformation phase. In 
\cite{Messenger:2011gi} a lower frequency cutoff of 10\,Hz was used; if 
this cutoff is reduced to 3\,Hz, improving the accuracy of parameter 
estimation, the errors on recovered redshift are reduced by tens of 
percent, up to nearly a factor 2 improvement for sources at redshift 4 
\cite{privateMessenger}. 

As the understanding of the achievable sensitivity for Einstein Telescope 
evolved during the Design Study, different sensitivity curves were 
considered. Early in the study the possibility was envisaged of pushing 
the low frequency limit down to 1--3\,Hz in a single interferometer, while 
still achieving a broad-band improvement of an order of magnitude in 
sensitivity over Advanced detectors \cite{ET-B}. However, this is highly 
challenging, and perhaps technically unfeasible, since the technology 
that achieves better high frequency ($\gsim 100$\,Hz) sensitivity, 
\emph{i.e.}\ higher laser power to bring down the photon shot noise, is in 
direct conflict with that required to improve low frequency ($\lsim 
100$\,Hz) sensitivity, \emph{i.e.}\ lower thermal noise and radiation 
pressure noise, since these are both worsened by higher laser power. 

Another design subsequently considered is the 
so-called \emph{xylophone} configuration \cite{ET-C}. The idea is to
install two interferometers in each V of the triangle, one that has good 
high-frequency sensitivity and the other with good low-frequency 
sensitivity \cite{ET-C,ET-D}. 
Several other new technologies, for instance frequency-dependent 
squeezing of light, have been studied in detail for the ET design~\cite{DSD} 
and must be combined to achieve the sensitivity goals of third 
generation detectors \cite{Hild:2011ub}. 

The main design parameters for ET to achieve a factor 10 improved 
sensitivity over advanced detectors, while also achieving good sensitivity 
in the 3--10\,Hz region, are as follows: 
10\,km arm lengths, 500\,W of input laser and 3\,MW of 
arm cavity power for the high frequency interferometer, and 3\,W of 
input laser and 18\,kW of arm cavity power and the use of cryogenic 
technology (mirrors cooled to 10\,K), for the low frequency interferometer 
\cite{ET-D,DSD}. Fig.~\ref{fig-ET}, right panel, compares the sensitivity 
of the initial single-interferometer configuration (ET-B) \cite{ET-B} with 
the xylophone configuration (ET-D) \cite{ET-D} which was the latest and 
most evolved design.\footnote{Note that the low-frequency sensitivity 
floor of ET-D, compared to ET-B, is determined by a more detailed and
realistic modelling of the suspension \cite{ET-D}.}
Also plotted for comparison are the design sensitivity curves of 
advanced LIGO (high power, zero detuning: `aLIGO') \cite{aLIGO} and 
advanced Virgo (`aVirgo') \cite{aVirgo}, and initial LIGO 
\cite{LIGODesign} and Virgo \cite{VirgoDesign}. 

\subsection{Response function and antenna pattern}
Let us begin by looking at ET's response to GW signals.
Far away from a source, gravitational waves emitted by
a system can be expressed in a suitable coordinate system
as a transverse and symmetric trace-free (STF) tensor
$h^{ij}$ (all temporal components of the metric perturbation 
vanish) whose only non-zero spatial components are 
\begin{equation}
 h^{11}=-h^{22}=h_+, \quad h^{12}=h^{21}=h_\times.
\end{equation}
Let $({\mathbf e}_x,\, {\mathbf e}_y,\, {\mathbf e}_z)$
be an orthonormal triad in which the metric 
perturbation takes the transverse-traceless form.
Then, using basis polarization tensors defined as
\begin{equation}
 {\mathbf e}_+ \equiv {\mathbf e}_x\otimes {\mathbf e}_x -
 {\mathbf e}_y \otimes {\mathbf e}_y, \quad 
 {\mathbf e}_\times \equiv {\mathbf e}_x\otimes {\mathbf e}_y
 +{\mathbf e}_y\otimes {\mathbf e}_x,
\end{equation}
the metric perturbation can be written as 
\begin{equation}
 {\mathbf h}=h_+\, {\mathbf e}_+ + h_\times\, {\mathbf e}_\times.
\end{equation}
ET's interferometers can also be represented as STF tensors: 
\begin{eqnarray}
 {\mathbf d}^1 & = & \frac{1}{2}({\mathbf e}_1 \otimes {\mathbf e}_1 - 
 {\mathbf e}_2 \otimes {\mathbf e}_2), \nonumber \\
 {\mathbf d}^2 & = & \frac{1}{2}({\mathbf e}_2 \otimes {\mathbf e}_2 - 
 {\mathbf e}_3 \otimes {\mathbf e}_3), \nonumber \\
 {\mathbf d}^3 & = & \frac{1}{2}({\mathbf e}_3 \otimes {\mathbf e}_3 - 
 {\mathbf e}_1 \otimes {\mathbf e}_1), 
\label{eq:det_tensors}
\end{eqnarray}
where ${\mathbf e}_1,$ ${\mathbf e}_2$ and ${\mathbf e}_3$ are unit vectors 
along the three arms of ET.  
The response $h^A(t),$ $A=1,2,3,$ of the interferometers to an 
incident gravitational wave is just the inner product of the 
detector tensor ${\mathbf d}^A$ with the wave tensor $\mathbf h$:
\begin{equation}
h^A(t) = d^A_{ij}\, h^{ij} 
     =  d^A_{ij} e_+^{ij}\, h_+ + d^A_{ij} e_\times^{ij}\, h_\times,
     \label{eq:response}
\end{equation}
which motivates definition of the antenna pattern functions $F^A_+$ and $F^A_\times:$
\begin{equation}
F^A_+\equiv d^A_{ij}\,e_+^{ij}, \quad
F^A_\times \equiv d^A_{ij}\,e_\times^{ij},
\end{equation}
in terms of which the response can be written as
\begin{equation}
h^A(t) = d^A_{ij}\, h^{ij} = F_+^A\, h_+ + F_\times^A \, h_\times.
\label{eq:response2}
\end{equation}

Let us now choose a coordinate system fixed to ET such that the three arms 
of ET's triangle are in the $xy$-plane and the unit vectors along the arms are
\begin{eqnarray}
{\mathbf e}_1 & = & \frac{1}{2} \begin{pmatrix} \sqrt 3, & -1, & 0\end{pmatrix},\quad
{\mathbf e}_2 = \frac{1}{2} \begin{pmatrix} \sqrt 3, & 1, & 0 \end{pmatrix},\nonumber\\
\mathbf{e}_3 & = & \begin{pmatrix} 0, & 1, & 0 \end{pmatrix}.\nonumber\\
\end{eqnarray}
Let $(\theta,\,\varphi)$ be the direction to the source in this
coordinate system with $({\mathbf e}_\theta,\, {\mathbf e}_\theta)$ 
denoting directions of increasing $\theta$ and $\varphi,$ respectively. 

The unit vectors ${\mathbf e}_x,$ ${\mathbf e}_y$
and ${\mathbf e}_z$ defining the radiation frame can be obtained by
successive counterclockwise rotations about the $z$-axis by an angle 
$\varphi$, about the new $y$-axis by an angle $\theta$ and the final 
$z$-axis by an angle $\psi$:
\begin{widetext}
\begin{eqnarray}
{\mathbf e}_x & = & (
- \sin\varphi\,\sin\psi + \cos\theta \,\cos\varphi\, \cos\psi,\, \,
  \cos\varphi\,\sin\psi + \cos\theta \,\sin\varphi\, \cos\psi,\,\,
-\sin\theta\,\cos\psi),\nonumber\\
{\mathbf e}_y & = & (
- \sin\varphi\,\cos\psi - \cos\theta \,\cos\varphi\, \sin\psi,\, \,
  \cos\varphi\,\cos\psi - \cos\theta \,\sin\varphi\, \sin\psi,\,  \,
-\sin\theta\,\sin\psi),\nonumber\\
{\mathbf e}_z & = & (\sin\theta \,\cos\varphi,\,\,
\sin\theta\,\sin\varphi,\,\, \cos\theta),\nonumber
\end{eqnarray}
where $\psi$ is the polarization angle defined by 
$\cos\psi = {\mathbf e}_\theta \cdot {\mathbf e_x}.$ 
The antenna pattern functions of the interferometer whose arms are 
$\left ({\mathbf e}_1,\,{\mathbf e}_2 \right)$ is:
\begin{eqnarray}
F^1_+ & = & 
-\frac{\sqrt{3}}{4} \bigl [(1+\cos^2\theta) \, \sin 2\varphi\, \cos 2\psi 
      +  2\, \cos\theta\,\cos2\varphi\,\sin 2\psi \bigr ], 
      \label{eq:f1p} \\
F^1_\times & = & 
+\frac{\sqrt{3}} {4}\bigl [ (1+\cos^2\theta) \, \sin 2\varphi\, \sin 2\psi  
      -  2\, \cos\theta\,\cos2\varphi\,\cos 2\psi \bigr ].
\label{eq:f1c}
\end{eqnarray}
\end{widetext}
The antenna pattern
functions are a factor $\sin\gamma=\sqrt{3}/2$ smaller than 
that of an L-shaped detector of the same length, where 
$\gamma=\pi/3$ is the opening angle of ET's interferometer arms.

The antenna pattern functions of the other two detectors in ET,
with arms $\left ({\mathbf e}_2,\,{\mathbf e}_3 \right)$ and
$\left ({\mathbf e}_3,\,{\mathbf e}_1 \right)$,
are obtained from $F_+^1$ and $F_\times^1$ by the transformation 
$\varphi\rightarrow \varphi \pm 2\pi/3:$ 
\begin{eqnarray}
F_{+,\times}^2(\theta,\varphi,\psi) & = & 
F^1_{+,\times}(\theta,\varphi+2\pi/3,\psi),\\
\label{eq:f2}
F_{+,\times}^3(\theta,\varphi,\psi) & = & 
F^1_{+,\times}(\theta,\varphi-2\pi/3,\psi).
\label{eq:f3}
\end{eqnarray}
$F^A_+$ and $F^A_\times$ are sometimes called antenna \emph{amplitude}
pattern functions to distinguish them from their squares ($F^A_+)^2$
and $(F^A_\times)^2,$ which are called antenna \emph{power} pattern 
functions \cite{Schutz:2011tw}.  The overall response of an interferometer to 
an incident wave depends on the square root of the sum of the 
antenna power pattern functions $F_+^2+F_\times^2.$ The joint 
response of all the three detectors in the ET network is 
\begin{equation}
F^2= \sum_{A=1}^3 (F_+^A)^2+ (F_\times^A)^2,
\end{equation}
which 
can be shown to be equal to
\begin{equation}
F^2=\frac{9}{32} \left ( 1+6\cos^2\theta+\cos^4\theta \right ).
\end{equation}
Thus, the joint antenna power pattern depends only on the colatitude
$\theta$ of the source. ET's response is smaller compared to an 
L-shaped interferometer by a factor $\sqrt{3}/2$ due to the $60^\circ$
opening angle, but its 3 detectors enhance its response by $\sqrt{3}$, 
leading to an overall factor of $3/2$. This is indeed what we find: $F(0)=3/2$.
The response averaged over $\theta$ is $\sqrt{\langle F^2 \rangle}=\sqrt{2/5}\,F(0)
\simeq 0.63\,F(0)$ and its minimum value is $F(\pi/2)=F(0)/\sqrt{8} \simeq 0.35\,F(0).$ 
With an average response 63\% of its optimum and a worst response 35\% of its 
optimum, and with no null directions, ET has virtually all-sky coverage.

\subsection{Null stream}
\label{sec:null_introduction}
It follows immediately from Eqs.\,(\ref{eq:det_tensors}), (\ref{eq:response}) that 
the sum of the individual responses $\sum_A h^A$ is identically equal to zero. 
The sum of the responses of any set of Michelson interferometers forming a closed 
path is zero and is called the \emph{null stream}. As we shall discuss later, such a 
null stream is an invaluable tool in data analysis. 

Two L-shaped detectors with arm lengths of 7.5\,km (and total length 
of 30\,km), rotated relative to each other by an angle $\pi/4$, are 
completely equivalent to ET in terms of their response and resolvability 
of polarizations. 
However, their response cannot be used to construct a null stream.

\subsection{Distance reach to compact binaries}
In 1986 Schutz showed \cite{Schutz:1986gp} that inspiralling binary
systems are standard candles whose (luminosity) distance can be 
measured from the observed gravitational wave signal, without the need 
to calibrate sources at different distances. Our detectors are able to 
measure both the apparent and absolute luminosity of the radiation, and 
hence to extract the luminosity distance of such a source: the magnitude 
of the gravitational wave strain gives the apparent luminosity but the rate 
at which the signal's frequency changes gives the absolute luminosity. 

For 
simplicity we shall consider a binary that is located at an optimal 
position on the sky (overhead with respect to the plane of ET) and 
optimally oriented (i.e.\ its angular momentum is along the line of 
sight). The discussion below holds good even when these assumptions are
dropped, but the measurement of the various angular parameters would
be essential in order to disentangle the distance. This would require a 
network of three or more detectors with long baselines to triangulate 
the source's position on the sky. We will also only consider the GW 
quadrupole amplitude in this discussion; higher-order corrections to the 
amplitude do not affect our conclusions on ET's distance reach.

The magnitude of the strain measured by our detectors when 
the signal frequency reaches the value $f$ is
\begin{equation}
h=\frac{4 \pi^{2/3} (G {\cal M})^{5/3}} {c^4 D} f(t)^{2/3}\,
\cos\left[\int_0^t f(t')\,{\rm d}t'\right],
\label{eq:amp}
\end{equation}
where $c$ is the speed of light, $G$ the gravitational constant, $\cal M$ is the chirp mass of the binary, related to its total mass 
$M=m_1+m_2$ and symmetric mass ratio $\nu=m_1m_2/M^2$ by ${\cal M}=
\nu^{3/5}M$, and $D$ is the proper distance to the source. Note that this 
expression is valid in the limit of asymptotically flat, static spacetime; 
we will soon discuss the effect of cosmological expansion on the observed 
signal. 

In addition to the signal's strain we can also measure the rate
at which its frequency changes\footnote{In reality we don't directly
measure the evolution of the frequency but use matched filtering to 
dig out the signal buried in noisy data. The end result, however, 
is the same. In fact, post-Newtonian approximation has allowed 
the computation of very accurate signal models which allows us to 
infer not only the chirp mass but also the mass ratio of the system.} via
\begin{eqnarray}
\frac{df}{dt} = \frac{96 \pi^{8/3}}{5} \left (\frac{G {\cal M}}{c^3}\right )^{5/3} f^{11/3}\\
 \Rightarrow \, 
{\cal M} = \frac{c^3}{G}  \left (\frac{5}{96 \pi^{8/3}} \frac{df}{dt} \right )^{3/5} f ^{-11/5}.
\label{eq:fdot}
\end{eqnarray}
Thus, measurement of the signal strain and rate of change of frequency 
can together determine the system's chirp mass and its distance from 
Earth.

For cosmological sources, however, the distance recovered by this 
method is not the comoving distance to the source $\chi$ (equivalent to
$D$ for a spatially flat FRW universe), but rather its luminosity distance 
$D_{\rm L} = (1+z)\chi$. This may be explained as follows: due to time 
dilation, the chirp mass of the system inferred from 
Eq.\ (\ref{eq:fdot}) 
will be ``redshifted" by a factor $(1+z)$, thus the signal will appear to have 
come from a source whose chirp mass is $(1+z){\cal M}$. 
%
Thus, if we reconstruct the masses of the binary from the frequency evolution 
of the waveform, we will obtain redshifted masses a factor $(1+z)$ larger than 
the physical masses of the system at redshift $z$. Symbols such as $m$, $M$, 
$\mathcal{M}$ will denote physical masses, whereas when discussing 
``redshifted'' observed mass parameters we will use a superscript $z$, for 
instance $m_1^z \equiv (1+z)m_1$.

This increase in apparent mass does not, however, mean that we will 
observe a greater signal amplitude: gravitational-wave amplitude, being 
dimensionless, cannot change due to redshift. Given this, and noting that 
${\cal M}f$ is invariant under the effect of redshift, we find that 
a source with physical chirp mass $\cal M$ will appear to us to have a 
chirp mass $(1+z){\cal M}$, and its apparent distance will be the luminosity 
distance $D_{\rm L} = (1+z)\chi$, instead of the proper or comoving distance. 


Let us now consider the distance reach of ET to an inspiral signal
from a compact binary of component masses $m_1$ and $m_2$, at a
luminosity distance $D_{\rm L}$ and whose orbit (assumed here to be 
quasi-circular) makes an angle $\iota$ with the line of sight.
There exist different measures of the distance reach of a detector: the 
\emph{horizon distance} is commonly used in data analysis (see, for 
instance, \cite{Abbott:2009tt}), while \emph{detector range} and 
\emph{range functions} were defined by Finn and Chernoff 
\cite{Finn:1992xs} and are routinely used as a measure of detector 
performance. Our measures of distance reach are inspired by all
of these concepts.

The signal-to-noise ratio (SNR) $\rho_A$ for a given signal (such as 
from an inspiralling binary), detected by matched filtering with an 
optimum filter, in a detector labelled $A$, is
\begin{equation}
 \rho_A^2 = 4 \int_0^\infty \frac{|H_A(f)|^2}{S_n(f)}\, {\rm d}f,
\end{equation}
where $H_A(f)$ is the Fourier transform 
of the response of detector $A$ and $S_n(f)$ is the one-sided noise power 
spectral density (PSD) of the detector, which we assume to be the same 
for all three detectors in the ET array.  A good analytical fit \cite{den}
to the ET-B noise PSD is given by $S_n(f) = 10^{-50} h_n(f)^2\,{\rm Hz}^{-1}$, where
\begin{eqnarray}
h_n(f) & = & 2.39\times 10^{-27}\, x^{-15.64} + 0.349\, x^{-2.145} \nonumber \\ 
       & + & 1.76\, x^{-0.12} + 0.409\, x^{1.10},
       \label{eq:psd}
\end{eqnarray}
and where $x=f/100\,\rm Hz.$ 
We may write the detector response in terms of
two GW polarizations via $H_A(f)=F_+^AH_+ + F_\times^A H_\times$, where
\begin{eqnarray}
H_+(f) & = & \sqrt\frac{5}{24}\, 
\frac{(G \mathcal{M}^z)^{5/6}}{\pi^{2/3} c^{3/2} D_{\rm L}}\, \frac{(1+\cos^2\iota)}{2}\, f^{-7/6},\\
H_\times(f) & = & \sqrt\frac{5}{24}\, 
\frac{(G \mathcal{M}^z)^{5/6}}{\pi^{2/3} c^{3/2} D_{\rm L}}\, \cos\iota\, f^{-7/6}.
 \label{eq:fourier}
\end{eqnarray}
The coherent SNR $\rho$ for the ET network, for uncorrelated noises in the 
three detectors, is simply the quadrature sum of the individual SNRs: $\rho^2=\sum\rho_A^2$. 
We discuss possible correlated noise in Sections \ref{sec:noise}, for the Gaussian 
noise budget, and \ref{sec:CBCfuture}, concerning possible correlated noise transients. 
For the present idealized sensitive range calculation we consider uncorrelated 
noises. 

For low mass systems such as BNS, the SNR is dominated by the inspiral part
of the signal; the coherent SNR can then be shown to reduce to
\begin{equation}
\rho^2 = \frac{5}{6}\, 
\frac{(G \mathcal{M}^z)^{5/3}{\cal F}^2}{c^3 \pi^{4/3}\,D^2_{\rm L}} 
\int_{f_1}^{f_2}\, 
\frac{f^{-7/3}}{S_n(f)}\,{\rm d}f,
\end{equation}
where $f_1$ and $f_2$ are lower and upper frequency cutoffs chosen so that 
the integral has negligible (say, $<1\%$) contribution outside this range and 
${\cal F}$ is a function of all the angles given by 
\begin{equation}
{\cal F}^2\equiv 
\sum_A \left [ \frac{1}{4} (1+\cos^2\iota)^2\, (F_+^A )^2
+ \cos^2\iota\, (F_\times^A  )^2 \right ].
\end{equation}
Here $F_{+\,\times}^A,$ $A=1,\,2,\,3,$ are the antenna pattern functions of the
detector given by Eqs.\,(\ref{eq:f1p})-(\ref{eq:f3}). Substituting for the antenna
pattern functions and summing over the three detectors gives
\begin{eqnarray}
{\cal F}^2(\theta,\,\varphi,\,\psi,\,\iota) & = & 
\frac{9}{128} \left (1+\cos^2\iota\right )^2\,\left (1+\cos^2\theta\right )^2 \cos^22\psi \nonumber\\
& + & \frac{9}{32} \left (1+\cos^2\iota\right )^2\,\cos^2\theta \sin^22\psi\nonumber\\
& + & \frac{9}{32} \cos^2\iota\,\left (1+\cos^2\theta\right )^2\sin^22\psi\nonumber\\
& + & \frac{9}{8} \cos^2\iota\,\cos^2\theta \cos^22\psi. 
\end{eqnarray}
The quantity $\cal F$ determines the SNR of a source of a given (observed) chirp 
mass at any given distance. Although the antenna power pattern $F^2$ is 
independent of $(\varphi,\,\psi),$ the quantity $\cal F$ is only independent of 
$\varphi.$ 
For certain source locations and orientations, the response is still independent of the 
polarization angle. For instance, either when the source is ``overhead" with respect 
to ET's plane (\emph{i.e.}\ $\theta=0,\pi$) or face-on (\emph{i.e.}\ $\iota=0,\pi$), 
$\cal F$ is independent of $\psi.$  It depends weakly on $\psi$ for values of $\theta$ 
and $\iota$ significantly different from these extreme values. The maximum value 
$\mathcal{F}_{\rm max} = 3/2$ is obtained when $\theta=\iota=0$, while the value 
of $\mathcal{F}^2$ averaged over $(\theta,\,\psi,\,\iota)$ is 
\begin{equation}
\overline{{\cal F}^2}= \frac{1}{8\pi}\int_0^\pi  \int_0^\pi \int_0^{2\pi} 
{\cal F}^2\,\sin\theta\,\sin\iota\,{\rm d}\theta\,{\rm d}\iota\,{\rm d}\psi
=\frac{9}{25}.\nonumber
\end{equation}
So the root-mean-square value of $\cal F$ is ${\cal F}_{\rm rms}\equiv
\sqrt{\overline{{\cal F}^2}}=3/5.$
%
%
The \emph{horizon distance} $\hat D_{\rm L}$ of a detector is defined as the maximal 
distance at which an optimally oriented, overhead binary (\emph{i.e.}\ $\iota=\theta=0$) 
can be detected above a threshold SNR of $\rho=\rho_T$, chosen large enough to keep the 
false alarm rate acceptably low; $\rho_T=8$ is considered reasonable for current detectors. 
Noting that ${\cal F}=3/2$ when $\iota=\theta=0,$ for ET the horizon is given by 
\begin{equation}
\hat D_{\rm L} \equiv  
\sqrt\frac{15}{8}\, 
\frac{(G \mathcal{M}^z)^{5/6}}{\pi^{2/3} c^{3/2} \,\rho_T}
\left [ \int_{f_1}^{f_2}\, 
\frac{f^{-7/3}}{S_h(f)}\,{\rm d}f \right ]^{1/2}.
\end{equation}
The horizon distance is not a very useful measure 
since essentially no signals can be detected beyond this distance with an SNR larger 
than $\rho_T.$ A more meaningful measure of the reach is the distance $\overline 
D_{\rm L}$ at which an ``average" source, meaning one for which 
${\cal F}={\cal F}_{\rm rms}=3/5$, produces an SNR of $\rho_T$. 
For such a source we obtain
\begin{equation}
\overline D_{\rm L} 
= \frac{3}{5}\, 
\hat D_{\rm L}.
\end{equation}
For a binary consisting of two components of (physical) mass $1.4\,M_\odot$ and 
for a threshold $\rho_{\rm T}=8$, we find $\overline D_{\rm L} \simeq 13\,\rm Gpc$ or 
$z=1.8$, and $\hat D_{\rm L}\simeq 37\,\rm Gpc$ or $z=4$; these distances can be larger 
for more massive binaries, and our simulated binary component masses extend up to 
$3\,M_\odot$. 
In our simulations, we inject signals of different orientations and polarization angles 
distributed uniformly over comoving volume up to a redshift of $z=6$. 

\subsection{Efficiency vs.\ distance}
The \emph{efficiency} of a detector at a given distance, and for binary sources
with given physical component masses, is the fraction of such sources 
for which ET achieves an expected SNR $\rho\ge \rho_T.$ 
ET will not be sensitive to every BNS merger at any given distance, but only 
to those that are preferentially located in certain sky directions and are 
suitably oriented \cite{Finn:1992xs}. 
The fraction $\epsilon(D_{\rm L})$ of sources detected by ET at a given luminosity
distance is given by
\begin{equation}
\epsilon(D_{\rm L}) = \frac{1}{8\pi} 
\int_0^\pi \int_0^\pi \int_0^{2\pi}
\Pi(\rho/\rho_T-1)\,\sin\theta\,\sin\iota\,{\rm d}\theta\,{\rm d}\iota\,{\rm d}\psi,
\label{eq:efficiency}
\end{equation}
where $\Pi$ is the unit step function $\Pi(x)=0$ if $x<0$ and $\Pi(x)=1$ if $x>0.$ 
Note that $\rho$ is a function of all angles, luminosity distance, redshift, etc. 
In Figure \ref{cbc_efficiency}, top right, we plot ET's efficiency as a function of 
redshift for binary neutron stars: the blue solid curve shows the efficiency for 
physical masses $m_1=m_2=1.4\,M_\odot$, choosing a SNR threshold $\rho_T =8$ 
and a lower frequency cutoff $f_1=1$\,Hz. 
As shown in this figure, ET should have 50\% efficiency at a redshift of $z\sim 1.3$, 
while its efficiency at $z=1.8$ (distance at which the angle-averaged SNR is 8) is 30\%.

\section{Simulation of ET Mock Data}
\begin{figure*}
\centering
\hspace*{-0.5cm}
\includegraphics[angle=0,width=0.49\textwidth]{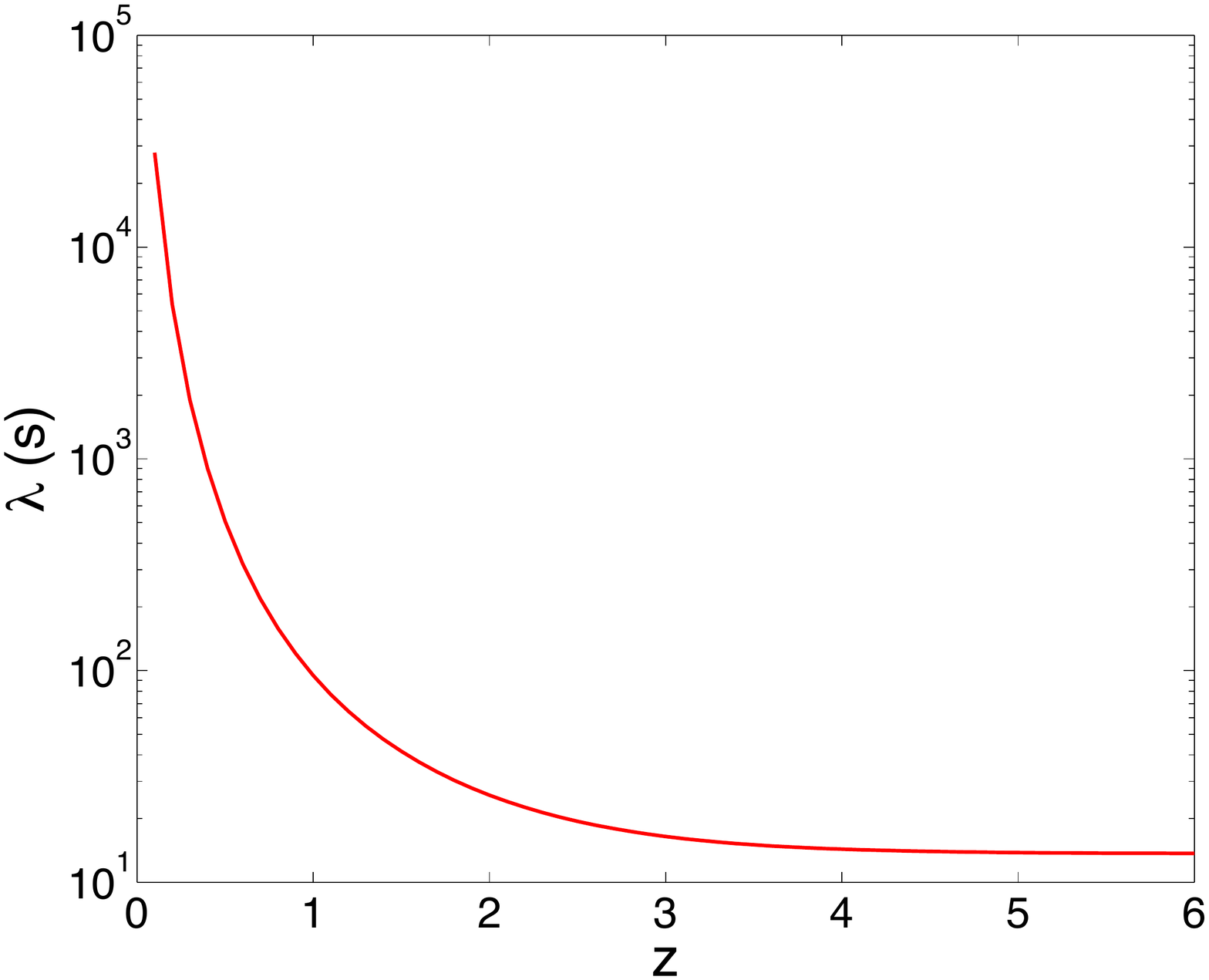}
\hspace*{0.1cm}
\includegraphics[angle=0,width=0.49\textwidth]{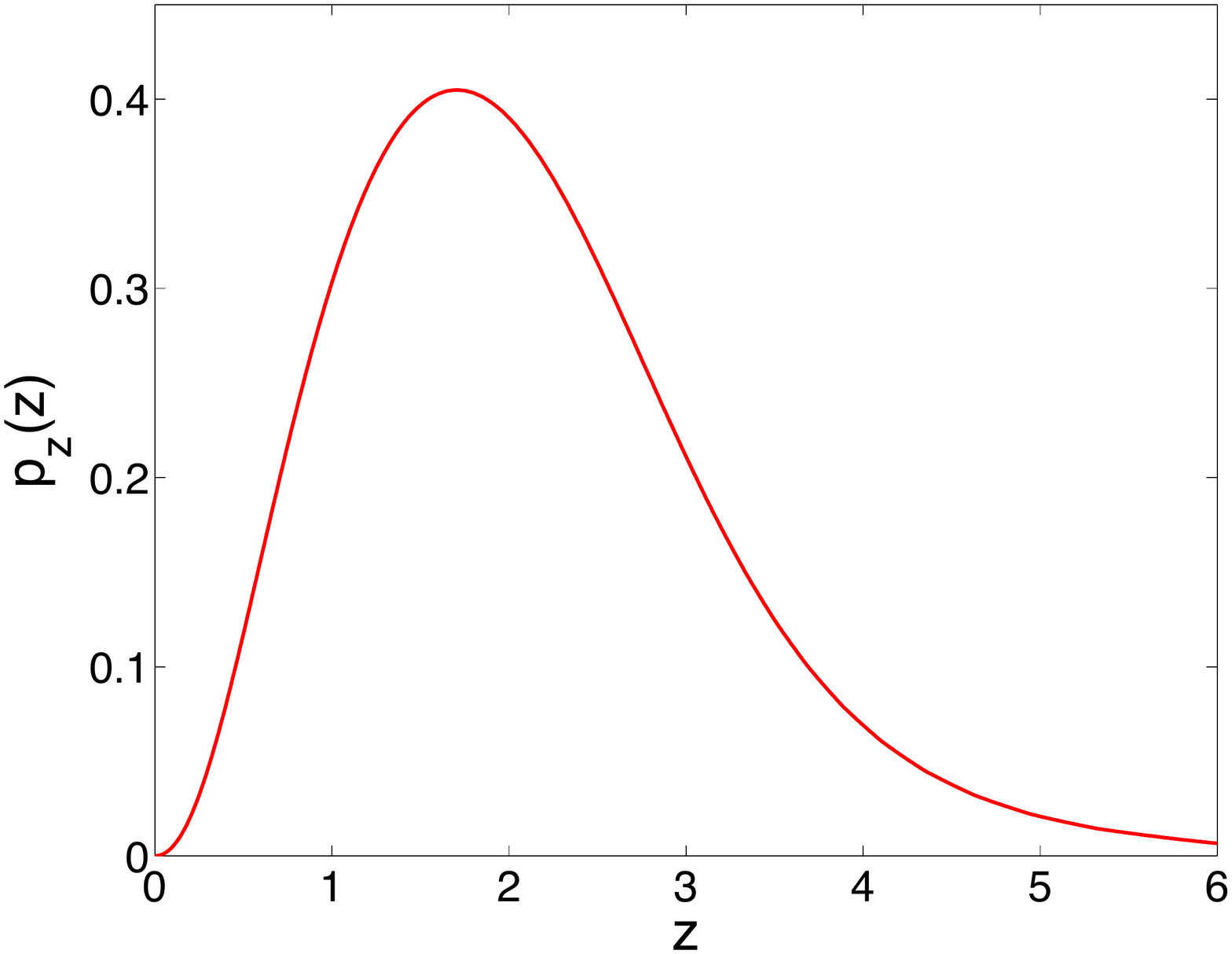}
\caption{Left: Average waiting time as a function of maximal redshift.
Right: Probability distribution of the redshift, assuming the star 
formation rate of \cite{hop06}, a distribution of the delay 
of the form $P(t_d) \propto 1/t_d$ with minimal delay of 20\,Myr and 
a local coalescence rate density of 1 Mpc$^{-3}$ Myr$^{-1}$.}
\label{fig-deltat}
\end{figure*}
In this section we will discuss how ET mock data was generated. We will describe
the cosmological model used and the rate of coalescence of binary neutron stars as
a function of redshift. We will also discuss how the background noise was generated
and the waveform model used in the simulation.

\subsection{Simulation of the GW Signal}
\begin{figure*}
\centering
\hspace*{-1.2cm}
\includegraphics[angle=0,width=1\textwidth,trim=0cm 3cm 0cm 3cm,clip]{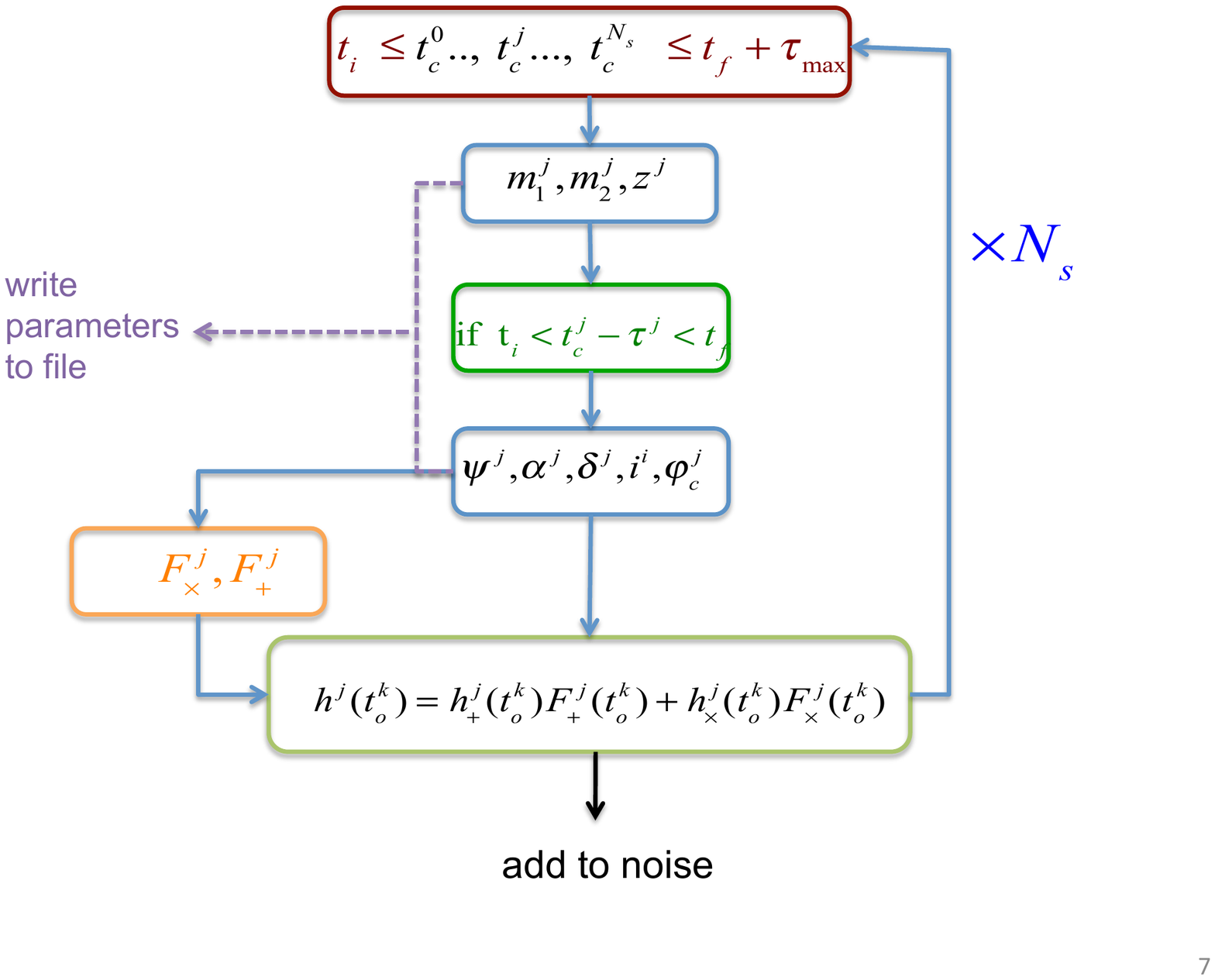}
\caption{Flow diagram of the Monte Carlo simulation code}
\label{fig-pipeline}
\end{figure*}

We use Monte Carlo techniques to generate simulated extra-galactic populations of binary 
neutron stars and produce time series of the gravitational wave signal in the frequency band of 
ET. We first describe how the distribution of injected BNS sources over redshift and mass was 
obtained, and then explain the simulation pipeline summarized in Fig.~\ref{fig-pipeline}. 

We first consider the rate of BNS coalescences in the Universe. We neglect the possible 
production of compact binaries through interactions in dense star systems, and we assume 
that the final merger of a compact binary occurs after two massive stars in a binary system 
have collapsed to form neutron stars and have inspiralled through the emission of gravitational 
waves. The merger rate tracks the star formation rate (SFR), albeit with some delay $t_d$ from 
formation of the binary to final merger. We use the SFR of \cite{hop06}, which is 
derived from new measurements of the galaxy luminosity function in the 
UV (SDSS, GALEX, COMBO17) and FIR wavelengths (Spitzer Space 
Telescope), and is normalized by the SuperKamiokande limit on the 
electron-antineutrino flux from past core-collapse supernovas.  This 
model is expected to be quite accurate up to $z \sim 2$, with very 
tight constraints at redshifts $z<1$ (to within $30-50 \%$). 

Following \cite{reg09}, we write the coalescence rate density $\dot{\rho}_c(z) $ (in 
Mpc$^{-3}$\,yr$^{-1}$) as: 
\begin{equation}
 \dot{\rho}_c(z) \propto \int_{t_d^{\min}}^{\infty} 
 \frac{\dot{\rho}_*(z_f(z,t_d))}{1+z_f(z,t_d)}P(t_d)\,{\rm d}t_d 
 \,\ \mathrm{with}\,\ \dot{\rho}_c(0)=\dot{\rho}_0\,,
\label{eq:rateV}
\end{equation}
where $\dot{\rho}_*$ is the SFR of \cite{hop06} (in M$_\odot$\,Mpc$^{-3}$ yr$^{-1}$), 
$z$ the redshift when the binary system merges, $z_f$ the redshift when the binary system 
is formed, $P(t_d)$ the probability distribution of the delay connecting $z$ and $z_f$, and 
$\dot{\rho}_0$ the rate density in our local universe. The normalization condition 
reproduces the local rate density for $z = 0$ and the factor $(1+z_f)^{-1}$ converts the 
rate density in the source frame into a rate density in the observer frame. 

The redshifts $z_f$ and $z$ are related by the \emph{delay time} $t_d$ which is the sum of 
the time from the initial binary formation to its evolution into a compact binary, plus the 
merging time $\tau_m$ after which emission of gravitational waves occurs. The delay is also 
the difference in lookback times between $z_f$ and $z$:
\begin{equation}
t_d = \frac{1}{H_0}
\int_z^{z_f} \frac{{\rm d}z'}{(1 + z')E(\Omega, z')}\,.
\end{equation}
where
\begin{equation}
E(\Omega,z)=\sqrt{\Omega_{\Lambda}+\Omega_{m}(1+z)^3}\,.
\end{equation}
In these simulations, we have assumed a flat Universe with  $\Omega_m=0.3$ and
$\Omega_{\Lambda}=0.7$ and Hubble parameter $H_0=70$\,km\, s$^{-1}$\,Mpc$^{-1}$.

We assume a distribution of the form $P(t_d) \propto 1/t_d$, as suggested by population 
synthesis \cite{popsynth}, with a minimal delay $t_d^{\min}= 20$\,Myr, corresponding 
roughly to the time it takes for massive binaries to evolve into two neutron stars \cite{bel}. 
This broad model accounts for the wide range of merger times observed in binary pulsars; 
it is also consistent with short gamma ray burst observations in both late and early type 
galaxies \cite{ber06}. 

The coalescence rate per redshift bin is then is given by 
\begin{equation}
\frac{dR}{dz}(z)= \dot{\rho}_c(z) \frac{dV}{dz}\,, 
\label{eq:rate}
\end{equation}
where $dV/dz$ is the comoving volume element:
\begin{equation}
\frac{dV}{dz}(z)=4 \pi \frac{c}{H_0} \frac{r(z)^2}{E(\Omega,z)}\,,
\end{equation}
where
\begin{equation}
r(z)= \frac{c}{H_0}\int_0^z \frac{{\rm d}z'}{E(\Omega, z')}\,,
\label{eq:distance}
\end{equation}
is the proper distance.

The average waiting time $\overline{\Delta t}$ between signals is calculated by taking 
the inverse of the coalescence rate, Eq.~(\ref{eq:rate}), integrated over all redshifts: 
\begin{equation}
\lambda = \left[\int_0^{z_{\max}} \frac{dR}{dz}(z)\,{\rm d}z\,\right]^{-1}.
\end{equation}
Fig.~\ref{fig-deltat}, left panel, shows $\Delta t$ as a function of the maximal redshift 
$z_{\max}$ out to which events are generated, given a local coalescence rate of 
$\dot{\rho}_0 = 1$\,Myr$^{-1}$\,Mpc$^{-3}$ which corresponds to the 
galactic rate estimated in \cite{kal04}, and which we adopt here.

We assume that signals arrive at the detector as a Poisson process and draw the
time intervals $\Delta t = t_c^{j+1}-t_c^j$ between successive coalescences at times 
$t_c^{j+1}$ and $t_c^j$, from an exponential distribution $P(\Delta t) = \exp(- \Delta t / \lambda)$. 
Coalescence times $t_c^k$ are generated between the start time of the observation $t_i$ 
and the end time $t_f$, to which we add the maximal duration $\tau_{\max}$ that a 
source can have in our frequency range (a 1.2\,+\,1.2\,M$_\odot$ system at $z=0$). 

Then, we proceed as follows for each source:
\begin{enumerate}
\item The physical masses of the two neutron stars are drawn from a Gaussian distribution 
with mean $1.4\,M_\odot$ and variance $0.5\,M_\odot$, and are restricted to the interval
 $[1.2,3]\,M_\odot$.
\item The redshift is drawn from a probability distribution $p(z)$ obtained by normalizing 
the coalescence rate $dR/dz$ in the interval $0$--$z_{\max}$:
\begin{equation}
p_z(z)= \lambda \, \frac{dR}{dz}(z)\,.
\end{equation}

Next we calculate the duration of the waveform in our frequency range: 
\begin{equation}
\: \tau \sim 5.4\, {\rm day} \left(\frac{\mathcal{M}^z}{1.22 M_\odot}\right)^{-5/3} 
\left(\frac{f_1}{1\,\mathrm{Hz}}\right)^{-8/3},
\label{eq-tau}
\end{equation}
where $f_1$ is the low-frequency cutoff of the detector; due to computational 
limitations in this initial study we take $f_1 = 10\,$Hz for the simulated signals. 

 \item For each source visible in our observation time-window $[t_i,t_f]$, the source's 
 location in the sky, its orientation, the polarization angle and the phase at the coalescence 
 are drawn from uniform distributions.

 \item The gravitational wave signal $h(t)=F_+(t) h_+(t)+ F_{\times} h_{\times}(t)$ of the 
source is calculated for each detector E1, E2 and E3 and for each observation time 
$t^k_o$ until the frequency reaches $f_1,$ and is added to the time series.
In these simulations, we have used so-called TaylorT4 waveforms \cite{Buonanno:2002fy}, 
up to 3.5 post-Newtonian order in phase $\phi(t)$ and only the most dominant lowest 
post-Newtonian order term in amplitude:
\begin{eqnarray}
h_+(t)& =& A(t)(1+\cos^2 \iota) \cos[\phi(t)] \\ 
h_{\times}(t)& =& 2A(t) \cos \iota \cos[\phi(t)]
\end{eqnarray}
where $\iota,$ as before, is the inclination angle of the binary with respect to the line of 
sight.  

The signal amplitude is then
\begin{eqnarray*}
A(t)  &\sim& 2 \times 10^{-21}  \left(\frac{1\,{\textrm{Mpc}}}{D_{\rm L}}\right) 
   \left( \frac{{\mathcal{M}}^z}{1.2\,M_\odot} \right)^{5/3}
    \left( \frac{f(t)}{100\,{\textrm{Hz}}} \right)^{2/3}, 
\end{eqnarray*}
where the luminosity distance $D_{\rm L}$ is in Mpc, $\mathcal{M}^z$ in M$_\odot$, and where 
$f(t)$ in Hz is the instantaneous gravitational-wave frequency (twice the binary's orbital 
frequency) which increases monotonically as the system shrinks and gets closer to 
merger. For a description of the TaylorT4 approximant and how it relates to other 
waveform approximants, see \cite{approximants} and references therein. 
\end{enumerate}

Theoretically, neutron stars could have maximum dimensionless spins 
$\chi=cJ/(GM^2),$ where $J$ is the star's angular momentum and $M$ its
mass, as large as 0.5 to 0.7, depending on the equation of state 
\cite{Salgado:1994}. These are moderately large spins and including
spin effects in our waveform model would be essential for unbiased and
accurate parameter estimation in real searches. However, in this 
exploratory work we neglect spins, as our main aim is to investigate the 
difficulty of discriminating overlapping signals. From an astrophysical 
point of view, neutron stars in coalescing binaries, such as the Hulse-Taylor
binary, have rather small spins of $\sim 6 \times 10^{-3},$ which 
will not significantly affect the phase evolution of the signal.

\begin{figure*}
\centering
\includegraphics[angle=0,width=0.46\textwidth]{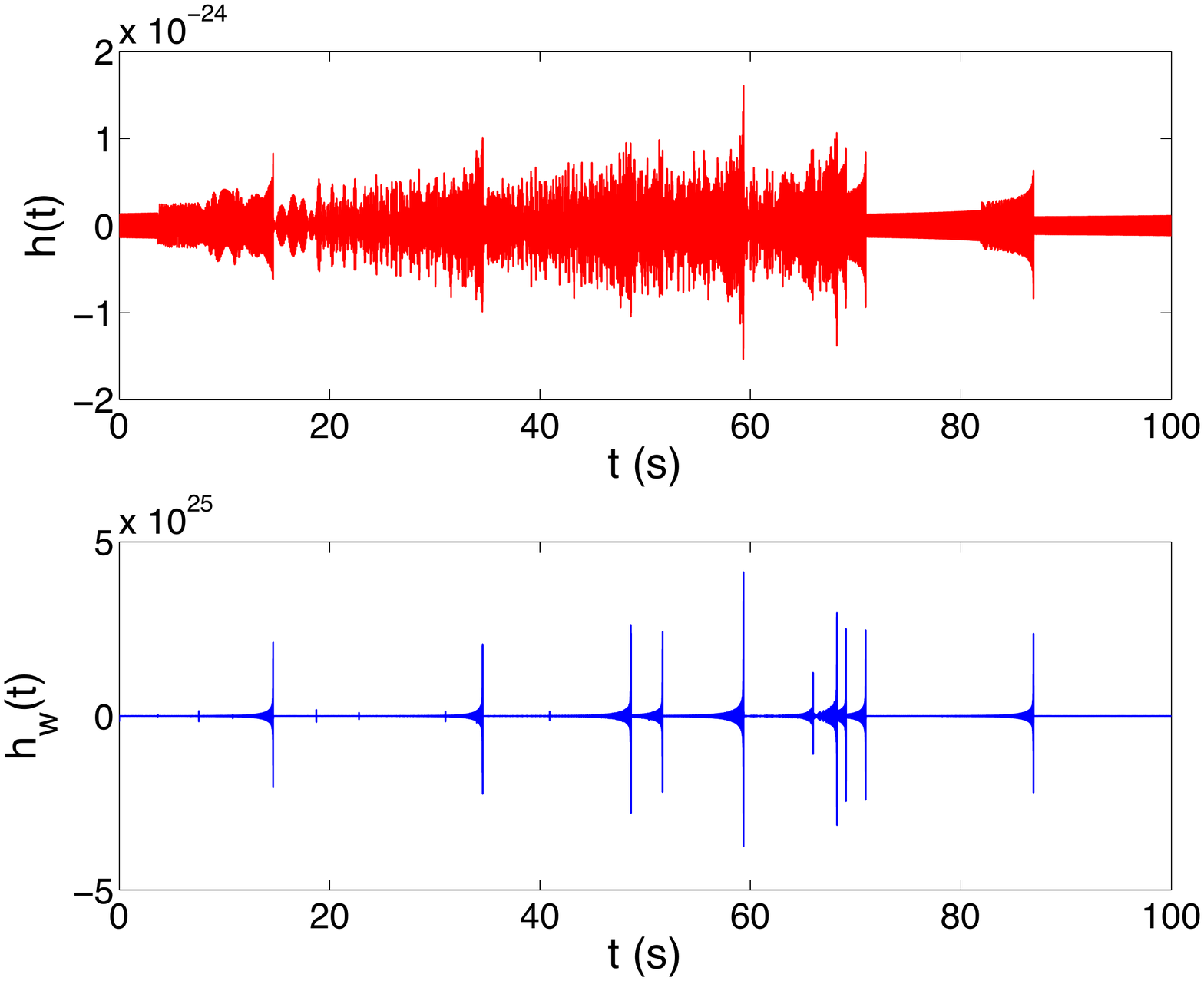}
\includegraphics[angle=0,width=0.49\textwidth]{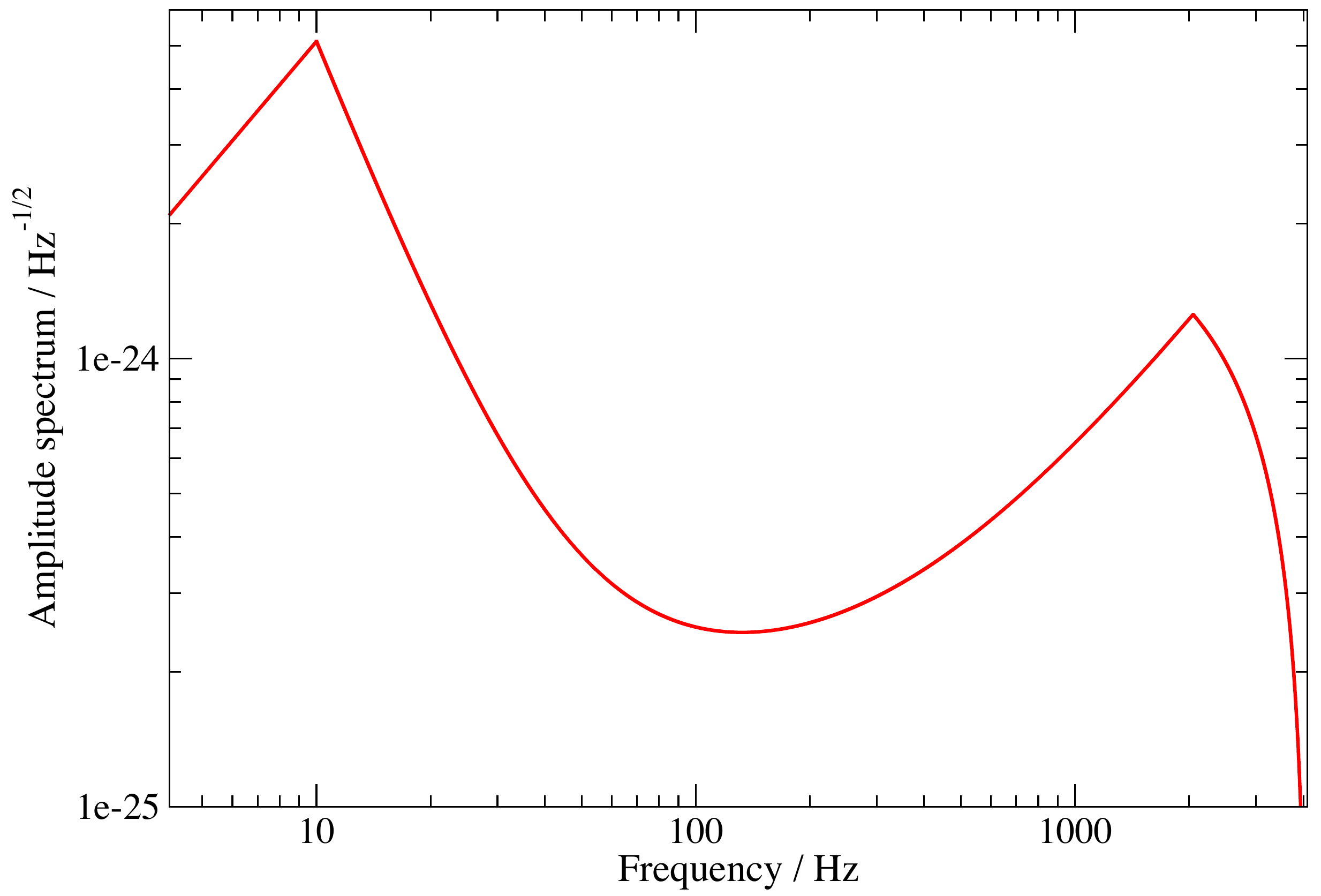}
\caption{
Left: Simulated time series of the gravitational strain at 
detector E1, for $z_{\max}=6$ and $f_1=10$\,Hz (top) and the 
same time series after the Fourier transform has been 
divided by the noise power spectral density of ET.
Right: The tapered projected ET noise spectrum used to color the 
noise.
Example audio files of the simulated GW signal alone or in the presence of noise can be 
found at the ET MDC website \url{http://www.oca.eu/regimbau/ET-MDC_web/ET-MDC.html}.}
\label{fig-seriesET}
\end{figure*}

An example time series of the gravitational wave signal including sources up to a redshift 
$z \sim 6$ (before adding simulated detector noise) is shown in the top plot of 
Fig.~\ref{fig-seriesET}, left panel. Although the sources overlap strongly in time, they 
are well separated in frequency, or become so when close to coalescence: an exception could 
be if two BNS signals with similar redshifted chirp masses were approaching coalescence within 
$<1\,$s of each other. Due to the form of the detector PSD, the main contribution to the 
matched filter power of any binary coalescence signal occurs when the chirping frequency is 
close to $100\,$Hz; the ``chirp'' is sufficiently rapid at (and after) this point that 
different sources can be clearly resolved. This is illustrated by the bottom plot of 
Fig.~\ref{fig-seriesET}, left panel, showing the optimal time domain filter, {\em i.e.}\ 
the inverse Fourier transform of the frequency-domain signal weighted by the noise power
spectral density. The detector PSD acts like a bandpass filter, weighting down
the lower frequencies where the signal spends most of its
time. The effective lenghts of signals, as ``seen" by ET, are, therefore, a lot
shorter than they actually are. Consequently, overlapping signals seen in the upper
panel do not lead to a loss in detection efficiency as we shall show in Sec IVB. 

\subsection{Simulation of the noise}
\label{sec:noise}

In order to produce the data set, it is necessary to use a model of the expected noise for 
the ET detectors. To this end, we assume that the noise will be stationary and Gaussian.
Moreover, for the time being, we assume that the noise realizations in the different 
detectors are uncorrelated. 

In reality the noise in collocated detectors will, most likely, have some correlated 
components, as has been seen in the two LIGO Hanford detectors. Different ET interferometers 
will be separately isolated in vacuum systems, thus we do not expect correlations of thermal 
or quantum noises, which form the main contribution above about $10\,$Hz. Common noise arises
most significantly as a result of having end stations with similar physical environments for 
two detectors: in particular, concerning ET's low-frequency sensitivity, seismic and Newtonian 
noise displacements. For ET, however, it is envisaged to stagger the end stations of the three interferometers by 1\,km. Whether this will reduce the common displacement noises to the extent 
that the detectors can be considered independent is a question under current investigation. 

The noise for each detector was generated using the following procedure: Firstly, we 
generate a Gaussian time series with a mean of zero, and unit variance. These time series 
are Fourier transformed and coloured by the relevant ET sensitivity curve in the frequency 
domain. To get the final time-domain representation of the noise, we apply an inverse 
Fourier transform.

The noise curve used is based on the analytic fit of Eq.~(\ref{eq:psd}) to the ET-B PSD 
discussed in Section II. 
To alleviate the effects of possible discontinuities across frame files, the PSD is 
gradually tapered to zero below the low frequency limit $f_l = 10$\,Hz, and 
above a frequency of $f_2 \equiv f_{\rm Nyquist}/2$. Fig.~\ref{fig-seriesET}, right panel,
shows the noise curve used to colour the data, with the tapering applied, for a sample 
rate of 8192\,Hz. The taper essentially acts like a bandpass filter and
removes power outside the band of interest. The absence of very high and very
low frequencies essentially assures continuity across the data segments
\cite{Frasca:1997}.

\section{First Analysis}

\subsection{Null stream}

A \emph{null stream} is a combination of the detector output streams such that the 
gravitational wave signal is identically zero and only 
noise remains. The existence of an ET null stream was noted already 
in~\cite{Freise:2008dk} and is a major motivation for the triangular triple Michelson
topology. 
Given an incident GW tensor $h_{ij}$, the three interferometer responses were derived in 
Eq.~(\ref{eq:response}), 
%
from which, as already remarked in Section~\ref{sec:null_introduction}, we find that
the sum of the three detector responses to any GW signal vanishes identically. 
We may define the \emph{null stream} as the sum of the strain time series $x(t)$ for the 
three ET detectors. For each single detector $A$ we have 
\begin{equation}
	x^{A}(t) \equiv n^{A}(t)+ d^{A}_{ij} h^{ij}(t), 
\end{equation}
where $n^{A}(t)$ is the noise realization, thus 
\begin{eqnarray}
x_{\rm null}(t) 	& \equiv & \sum_{A=1}^3 x^{A}(t) \nonumber \\
			& = & \sum_{A=1}^3 n^{A}(t) + \sum_{I=1}^3 d^{A}_{ij} h^{ij}(t) \nonumber \\
			& = & \sum_{A=1}^3 n^{A}(t)
\end{eqnarray}
is free of GW signals, and will also not contain any common (correlated) noise for which the 
sum over the three detectors happens to vanish.

If the noise properties are homogeneous among the detectors, 
\begin{equation}
S_n^{1}(f) \simeq S_n^{2}(f) \simeq S_n^{3}(f)\, ,
\end{equation} 
and if correlations between detectors can be neglected, we can use the 
null stream to estimate the average PSD in each of the three detectors. 
In this case,
\begin{eqnarray}
    \langle X_{\rm null}(f) X_{\rm null}^*(f') \rangle & = & \left< \sum_{A, B} N^{A} N^{B*} \right> \nonumber \\
& \simeq & \left< \sum_A N^{A} N^{A*} \right> \nonumber \\
& \simeq & 3 \frac{1}{2} \delta(f-f') \hat{S}_n^{A}(f),
\end{eqnarray}
where $X_{\mathrm{null}}(f)$ is the Fourier transform of $x_{\rm null}(t)$ and, in the last line, 
$\hat{S}_n^{A}(f)$ is the noise PSD in any of the three interferometers 
\emph{in the absence of a GW signal}. Defining
\begin{equation}
\left< X_{\rm null}(f) X_{\rm null}^*(f') \right> = \frac{1}{2} \delta(f-f') S_{n,\rm null}(f),
\end{equation}
we find 
\begin{equation} \label{eq:nullpsd}
	\hat{S}_n^{A}(f) \simeq \frac{1}{3} S_{n,\rm null}(f)
\end{equation}
as an estimate for the individual single-interferometer PSDs with the signals removed.

\begin{figure*}
	\resizebox{1.03\columnwidth}{!}{\includegraphics{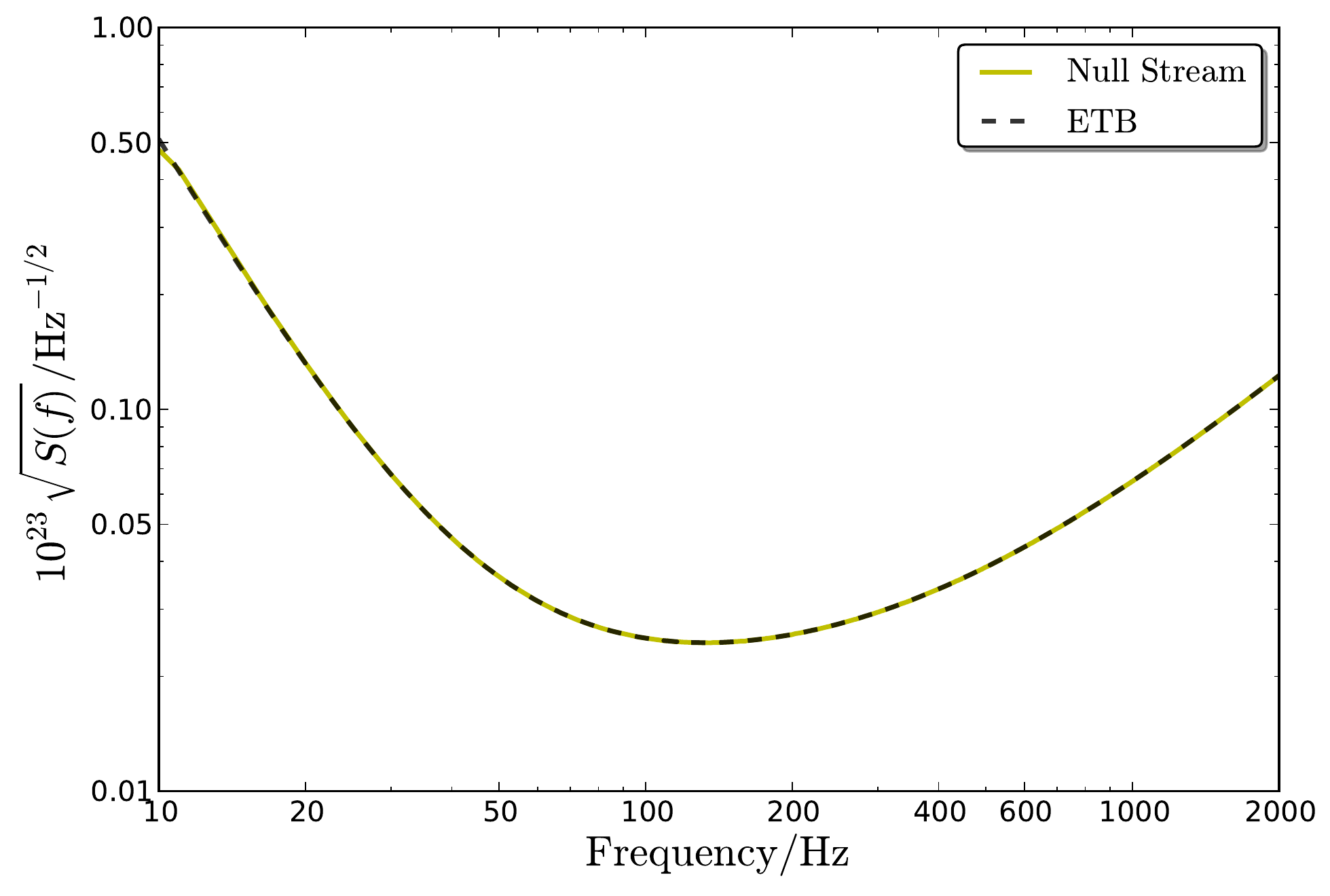}}
	\resizebox{\columnwidth}{!}{\includegraphics{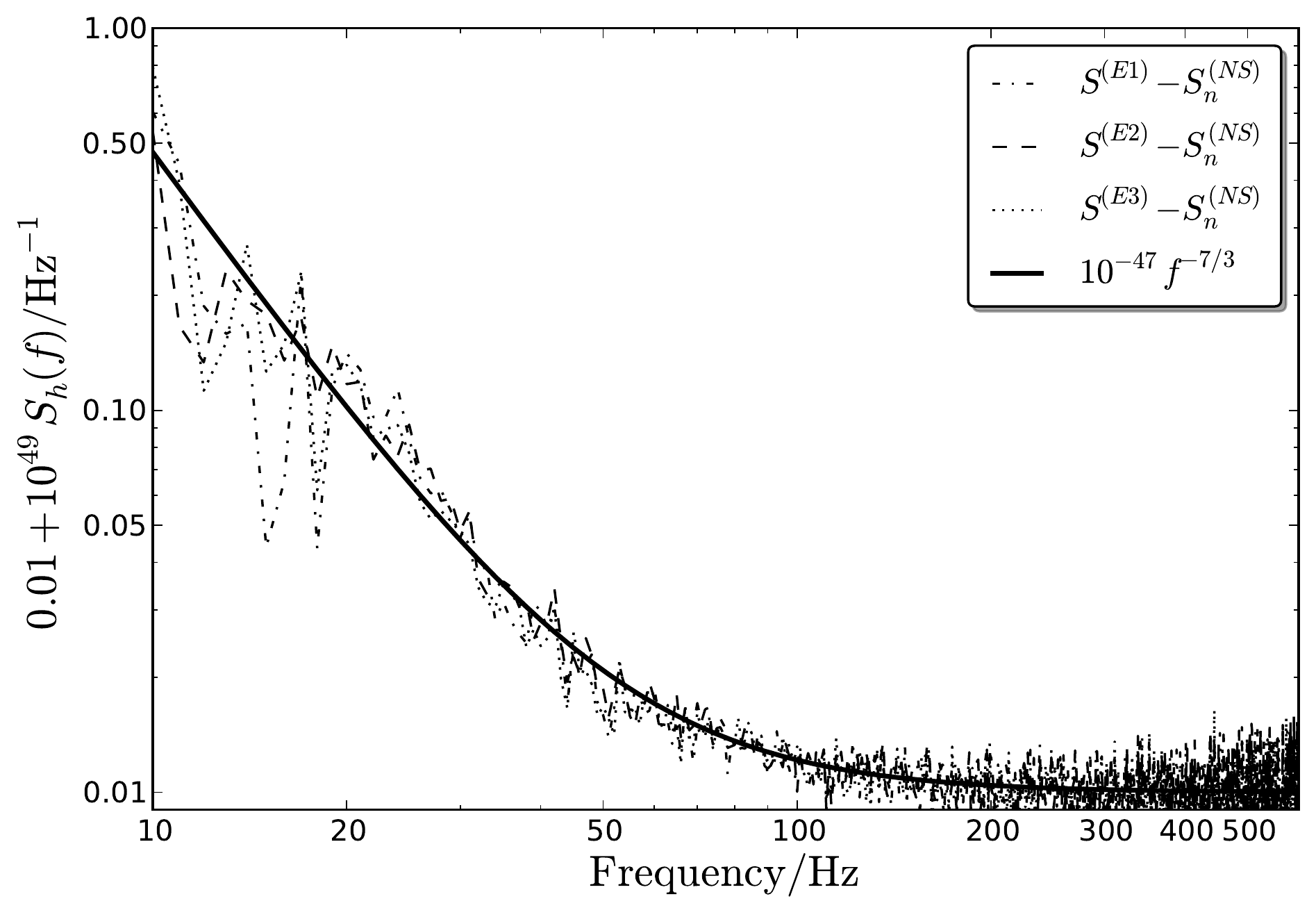}}
	\caption{\emph{Left}---Sample PSD calculated from the null stream, compared to the 
	theoretical ETB fit \ref{eq:psd}. The null PSD is estimated within data segments of length 
	2048\,s by averaging non-overlapping samples each of $1$\,s length, and is then 
	averaged over all 2048\,s long segments in the dataset. The figure shows $\sqrt{S_{n,\rm null}(f)/3}$.
	\emph{Right}---Difference between (one-third of) the null stream PSD and the 
	E$_j$ PSDs obtained by averaging over the whole dataset, as defined in Eq.~(\ref{h-hat}). 
	The residuals are consistent with the $f^{-7/3}$ spectrum expected from binary inspiral 
	signals. To aid visibility, the quantities plotted have been scaled by $10^{49}$ and the 
	constant 0.01 has been added.}
	\label{fig-nullstreamPSD}
\end{figure*}
The null stream PSD, which we plot on the left panel of Fig.~\ref{fig-nullstreamPSD}, 
then has the advantage of giving a better representation of the noise content of the 
three detectors. The typical sensitivity improvement is nonetheless fairly small, 
about 1\% in the 10 -- 100\,Hz band.  As a proof of principle of the effectiveness of 
the use of null stream PSD instead of the single detector one, we computed the median 
over the whole dataset of the difference between (one-third of) the null stream PSD and 
the individual detector PSD's $S_n^A(f)$. 
These residuals should be consistent with the median PSD of the injected signals in each detector:
\begin{equation} \label{h-hat}
S_n^A(f) - \frac{1}{3} S_{n,\rm null} \simeq \hat{H}(f)
\end{equation}
where 
$\hat{H}(f)$ is the power spectral density of GW signals. The result of this operation is 
shown in Fig.~\ref{fig-nullstreamPSD}, right panel. The residual spectrum between 10 and 
400\,Hz in each detector is consistent with the theoretical expectation $S_h(f)\sim f^{-7/3}$.


\subsection{Compact Binary Coalescence analysis}
We analysed the triple coincident simulated data using a modified version of the LIGO-Virgo 
Ihope pipeline \cite{Allen:2005fk,Brown:2005zs,Abbott:2009tt,Abadie:2010yb}
which is used to search for signals from compact binary coalescences (CBC). This pipeline
is a \emph{coincident} analysis: data streams from different detectors are separately
filtered against template waveforms and the resulting maxima of SNR are checked for 
consistency between detectors. The main motivation of this procedure is to reduce 
computational cost when analyzing data from spatially separated detectors with a-priori 
unknown duty cycles. 

Coherent analysis, where data streams are combined before finding maxima of SNR, should in 
principle be more sensitive at fixed false alarm rate if many detectors are involved 
\cite{Pai:2000zt,Pai:2001cf,Bose:2002by,Harry:2010fr,Bose:2011km}. For ET, the detector 
outputs could be combined into a null stream and synthetic $+$ and $\times$ detectors, and 
for the 2 non-null streams, the coherent detection statistic is then identical to the 
coincident one \cite{Harry:2010fr}. Hence unless other sites contribute there is no gain 
expected specifically from using a coherent analysis. In our case we might expect a small gain 
in sensitivity by using synthetic $+$ and $\times$ data, since it eliminates a fraction of the 
noise from each detector (the contribution to the null stream) while keeping all the signal 
power.\footnote{Alternatively, one can view this recombination as creating two synthetic 
detectors with $90^\circ$ opening angles and slightly better sensitivities than each of the
original three detectors.} However to establish an initial benchmark we have kept the existing 
framework where each physical detector is filtered separately. 

The stages of the coincident pipeline are as follows:
\begin{itemize}
\item Estimation of the PSD by median over several overlapping time chunks within a 2048\,s 
segment. We use the single-detector outputs rather than the null stream to estimate the noise: 
in principle loud signals could bias this estimation, however as shown in 
Fig.~\ref{fig-nullstreamPSD} any such bias is on average extremely small; we also compared
the estimated sensitivity 
over different segments of single-detector data 
and found negligibly small differences. 
\item Generation of a template bank covering the chosen parameter space of binary masses
\item Matched filtering of each template against the data stream of each detector to generate 
an SNR time series $\rho(t)$
\item Trigger generation: for each template, maxima of SNR over a sliding time window of 
length 15\,s were found, and a ``trigger'' was generated if any such maxima exceeded an SNR of 
$5.5$
\item Clustering to reduce trigger numbers: if there are multiple triggers within a small 
region of parameter space (binary masses plus time \cite{trigscan_poster}) the trigger with 
largest SNR is selected and others in the region are discarded
\item Coincidence between detectors: only pairs or triples of triggers with consistent 
coalescence times and masses \cite{ethinca} survive and are designated as events. 
\item Ranking of events by combined SNR$^2$, $\rho_{\rm C}^2$ (sum of $\rho^2$ over coincident 
triggers).
\end{itemize}
There are several differences compared to standard LIGO-Virgo searches. The main ones 
concern the frequency range of data searched, the parameter space of the search and the 
method for determining the significance of candidate events. 

The length of an inspiral template increases rapidly with the lowest frequency that is 
matched filtered in the analysis (Eq.~\eqref{eq-tau}). For technical reasons related to 
memory load and PSD estimation, the standard matched filter code used for LSC-Virgo analyses 
\cite{Allen:2005fk} cannot filter templates longer than a few minutes: hence we chose to 
impose a lower frequency cutoff of 25\,Hz. This limitation should be addressed in future 
analyses, and may be relevant to analysis of Advanced LIGO/Virgo data.

The template bank was chosen to cover the possible range of redshifted (\emph{i.e.}\ observed) 
mass pairs corresponding to the BNS injections up to redshift 4. The minimum component mass 
was taken as $1.2\,M_\odot$; with a maximum injected component mass of $3\,M_\odot$, the 
observed total mass at $z=4$ is then $15\,M_\odot$, which we took as our maximum component 
mass, with a maximum total mass of $30\,M_\odot$. The maximum injected mass ratio is $3/1.2 = 
2.5$ corresponding to a ``symmetric mass ratio'' $\eta = m_1m_2/(m_1+m_2)^2 \simeq 0.204$, 
thus templates with $\eta<0.2$ were removed, considerably reducing the size of the bank. 


Since the simulated noise was Gaussian, signal-based vetoes and data quality vetoes were 
not necessary to suppress detector artefacts, and events were ranked simply by the quadrature
sum of SNR over coincident triggers. 
The noise background in our mock data is expected to be a function of combined SNR alone,
thus we set a threshold in the value of $\rho_{\rm C}$ above which we consider an event likely to 
be a true GW signal. 

Note that the time shift method used to estimate background event rates in LIGO-Virgo searches 
fails here. In order for such methods to be valid, the number of detectable GW events over the 
search time should be small (of order 1): otherwise, loud triggers due to true GW signals may 
significantly distort the background distribution, by forming random time-shifted coincidences 
with noise triggers. In the present case we see tens of thousands of detectable signals, 
thus the distribution of loud time-shifted coincidences is totally dominated by such 
`signal-plus-noise' events.

\subsubsection{Events found by CBC analysis}

\begin{figure}
  \centering
  \vspace*{-0.4cm}
  \includegraphics[width=1.0\columnwidth]{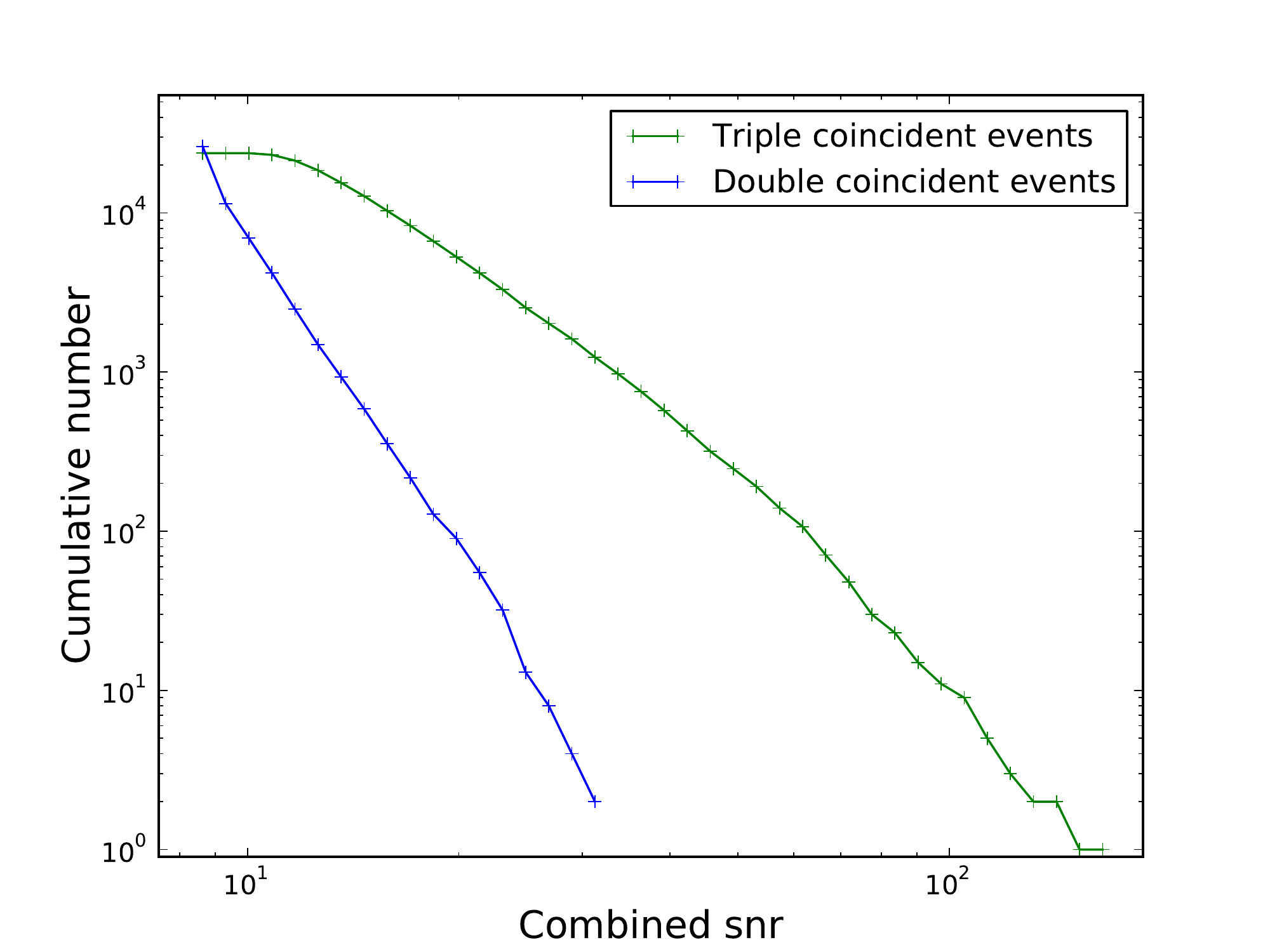}
  \caption{Cumulative histogram of CBC events as a function of combined SNR $\rho_{\rm C}$, 
  divided into double (two-detector) and triple (three-detector) coincidences.}
  \label{cbc_events_hist}
  \end{figure}

The CBC analysis outputs a list of loudest events with the coalescence time, combined SNR, 
and the component masses of the best-fitting template for each event. The distribution 
over $\rho_{\rm C}$ is plotted in Figure~\ref{cbc_events_hist} for both double- and 
triple-coincident events.
Given the single-detector SNR threshold $\rho_t = 5.5$, 
the quietest possible double coincidence has $\rho_{\rm C} = \sqrt{2}\rho_t \simeq 7.78$ and the 
quietest triple has $\rho_{\rm C} = \sqrt{3}\rho_t \simeq 9.53$.
The expected cumulative distribution of events from an astrophysical population is approximately
proportional to the inverse cube of combined SNR (thus to the cube of the luminosity distance, or 
to the volume of space seen by the search). Deviations from 
this inverse-cube behaviour will arise due to evolution of the source population over redshift, also 
because the physical volume of space is no longer exactly proportional to distance$^3$ at large 
$z$, and also since the observed masses of a coalescing binary are larger than the physical 
masses by a factor $(1+z)$, changing the expected SNRs. 

Over most of the range of $\rho_{\rm C}$ the distribution of triple coincidences is close to $\sim\rho_{\rm C}^{-3}$ 
as expected; 
with decreasing combined SNR values, 
an increasing fraction of signals are seen as double coincidences. We see no significant 
background distribution of triples, which would be expected to rise exponentially at small combined 
SNR. Thus in principle the efficiency of the search could be improved by lowering the SNR 
threshold. 

The distribution of double coincidences shows two components: an approximate power-law at 
higher $\rho_{\rm C}$ and a more rapidly rising component below about $\rho_{\rm C}=9$. We interpret 
these as a cosmological population of sources, modulated by the variation in the proportion 
found as doubles vs.\ triples; and a Gaussian noise background, respectively. Thus we expect 
that above a combined SNR $\rho_{\rm C}\gtrsim9$ the great majority of events will be caused by 
binary coalescence signals rather than random noise.

\subsubsection{Efficiency and accuracy}

We evaluate the search efficiency as a function of redshift by testing time coincidence between 
simulated signals 
and found events (using a ``coalescence time'' at which the chirping signal reaches a 
well-defined frequency) and choose a time window of $\pm 30$\,ms. For a given event or injection 
there are the following cases:
\begin{itemize}
\item False event: an event which does not fall within 30\,ms of an injection
\item True event: an event falling within 30\,ms of one or more injections
\item Missed injection: a simulated signal which does not fall within 30\,ms of an event
\item Found injection: an simulated signal within 30\,ms of a found event.\,\footnote{Note that 
events are clustered over time windows of a few seconds, thus more than one event cannot
be found within a 30\,ms window.}
\end{itemize}
However, if we have very frequent candidate events or injections, we may encounter significant 
numbers of
\emph{wrongly found injections}, meaning chance time coincidences between injections and 
noise events where the expected SNR of the injected signal is below the analysis threshold. 
For these we do not expect the estimated mass parameters and effective luminosity distance 
from the analysis pipeline to correspond to those of the simulated source; the fractional error in 
these parameters will be order(1). 
Wrongly found injections would lead us to overestimate the search efficiency and 
would degrade the accuracy of recovered source parameters. 
%
To minimize such effects 
whenever two or more simulated signals fall within $\pm30$\,ms of an event, we consider only 
the injection with the lowest redshift to be found. In practice this ambiguity is found to 
affect only a small fraction (sub-percent) of signals. In order to minimize possible bias in 
assessing the accuracy of recovered source parameters, we do not impose any further requirement 
(for instance, on the chirp mass) in order for an injection to be counted as found. 
\begin{figure*}
  \hskip -0.3cm
  \includegraphics[width=0.45\textwidth]{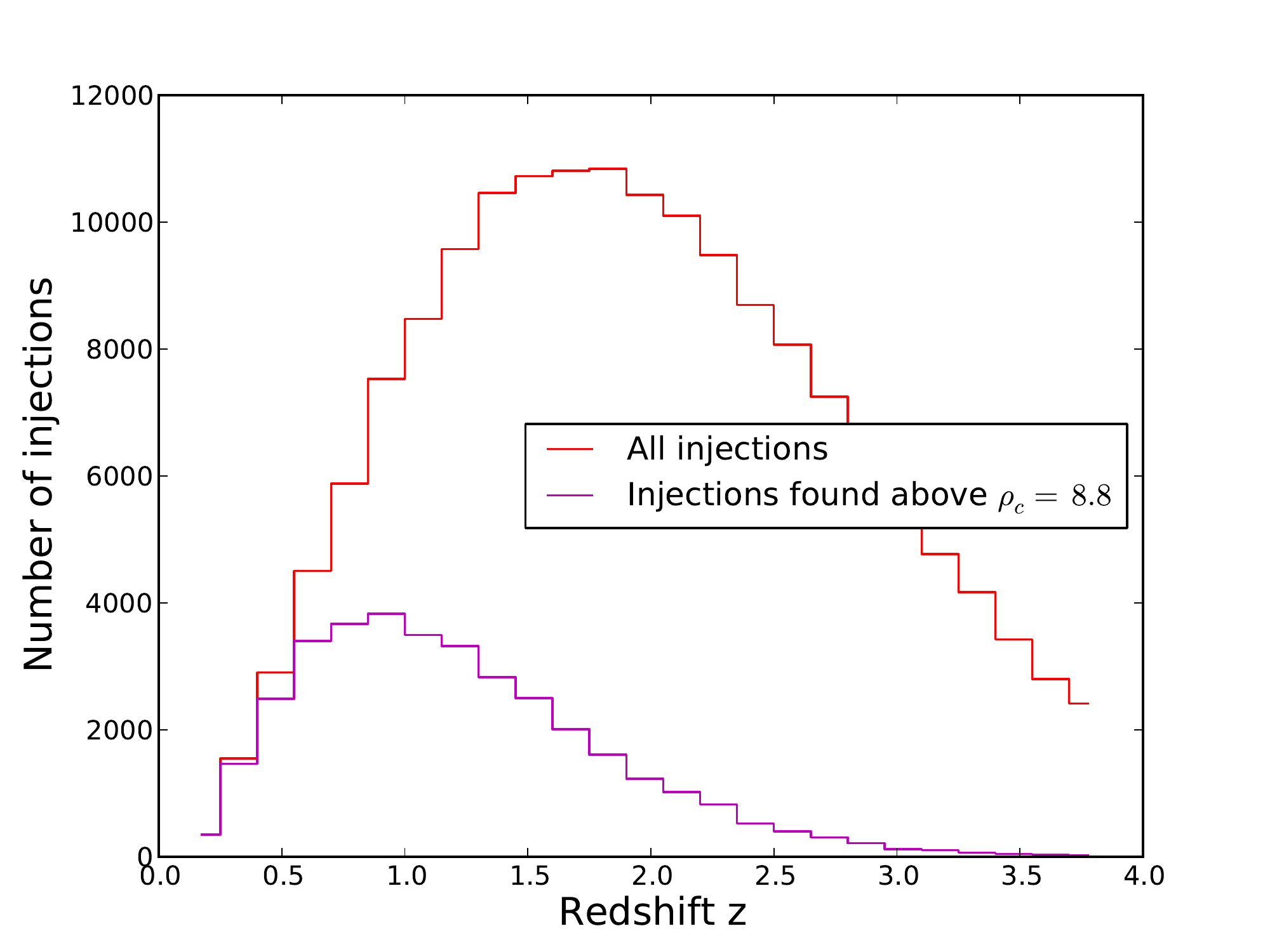}
  \hskip -0.5cm
  \includegraphics[width=0.45\textwidth]{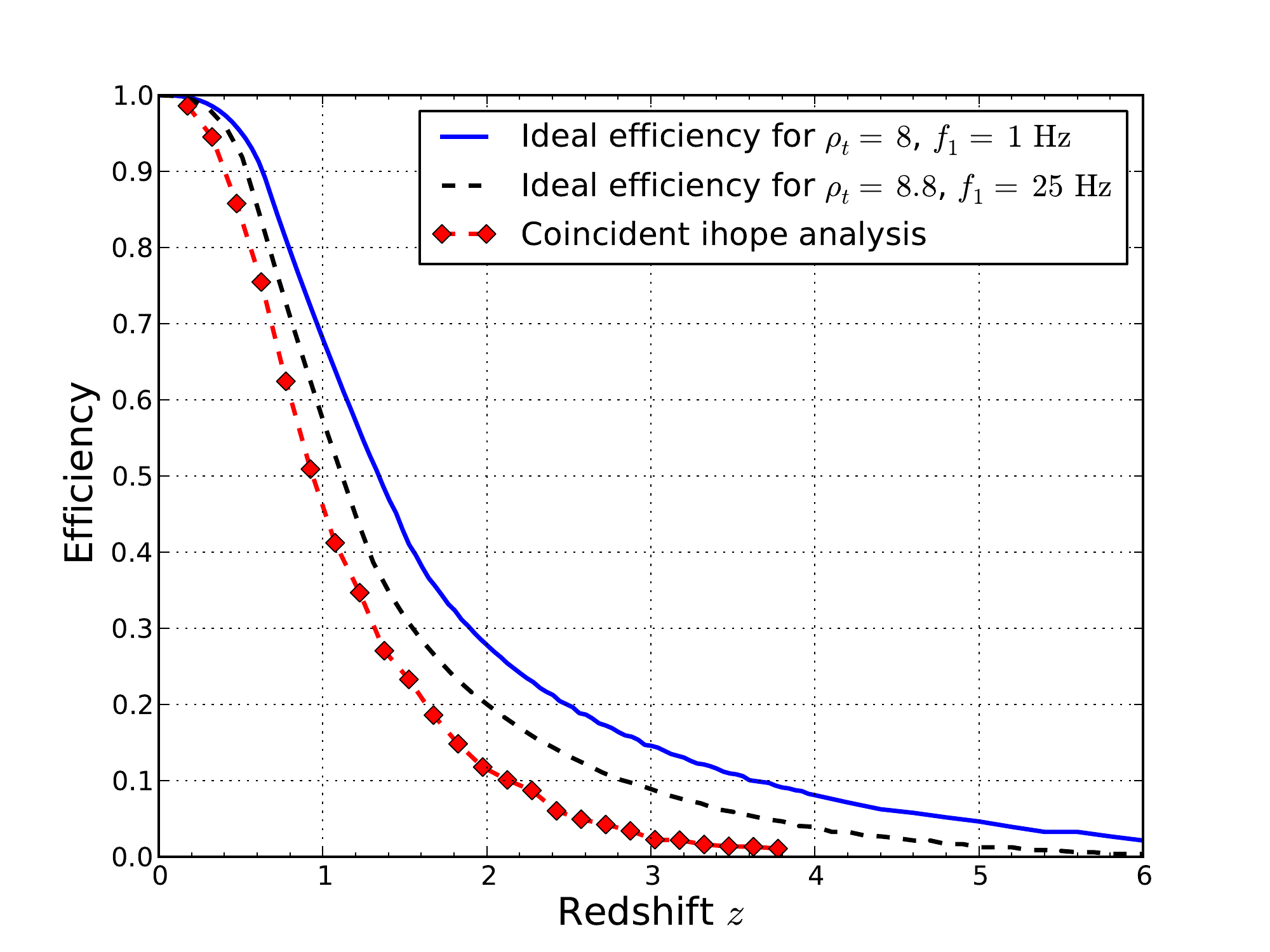} \\
  \hskip -0.3cm
  \includegraphics[width=0.45\textwidth]{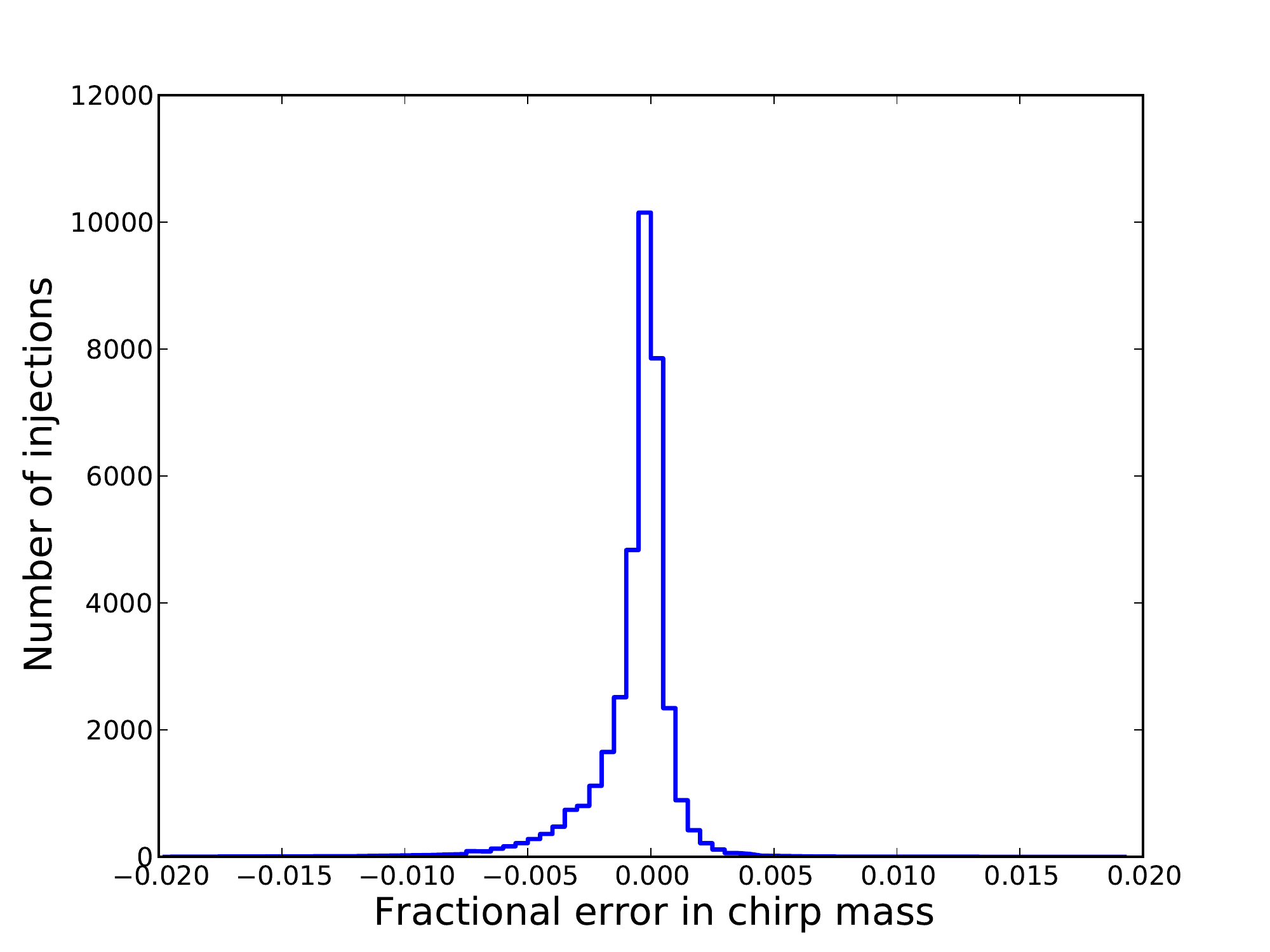}
  \hskip -0.5cm
  \includegraphics[width=0.45\textwidth]{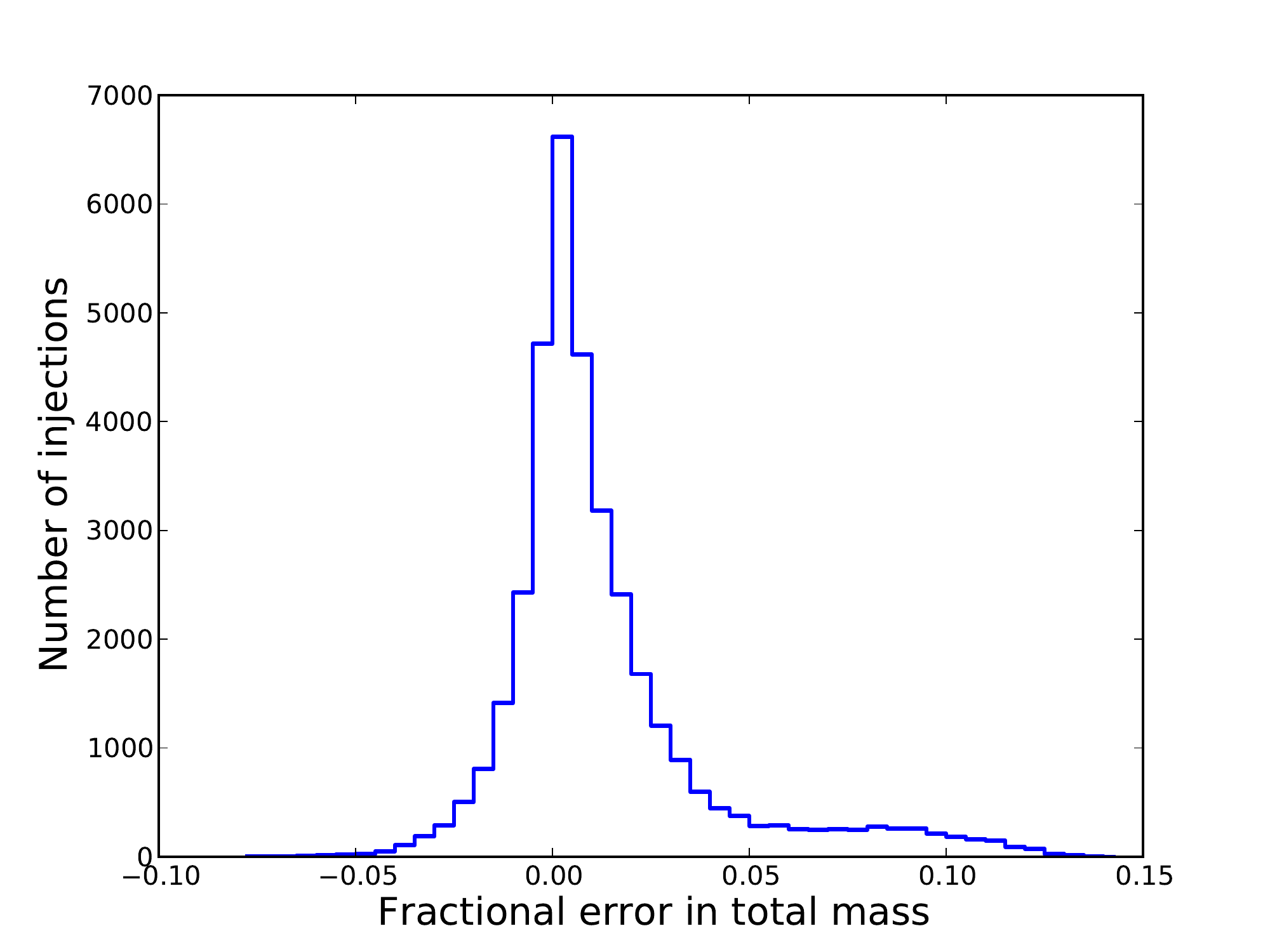}
  \caption{\emph{Top left}---Distributions of all BNS injections, and those found by the CBC
  pipeline, vs.\ redshift. Here events with $\rho_{\rm C}>8.8$ were considered as candidate signals.
  \emph{Top right}---Efficiency of the CBC search vs.\ redshift. We show the theoretical (ideal)
  efficiency as defined in Eq.~(\ref{eq:efficiency}) for a threshold SNR of $\rho_T=8$ and a low
  frequency cutoff $f_1=1$\,Hz, and also for $\rho_T=8.8$, $f_1=25$\,Hz for comparison with
  the signals found by the ihope pipeline.  
  \emph{Bottom left}---Histogram of fractional errors in chirp mass.
  \emph{Bottom right}---Histogram of fractional errors in total mass.}
  \label{cbc_efficiency}
\end{figure*}

As mentioned above, we require a threshold on $\rho_{\rm C}$ to limit the number of false
events caused by noise. Here we choose to impose $\rho_{\rm C}>8.8$, finding 36774 events 
above this threshold in the 2419200\,s of data analyzed. 
By comparing these with the catalogue containing 177350 simulated coalescence signals 
over the analysis time we find 850 false events, giving a directly estimated false alarm 
probability (FAP) of 2.3\%. The efficiency of finding injections as a function of redshift is 
summarized in Figure~\ref{cbc_efficiency}, top two panels. 

In the top right plot we compare the efficiency of the current analysis with the theoretical
ideal efficiency defined in Eq.~(\ref{eq:efficiency}), for two different values of the threshold 
SNR $\rho_T$ and the low frequency cutoff $f_1$. The ihope analysis does somewhat 
worse than the corresponding theoretical curve, which can in part be attributed to the 
single-detector SNR threshold $\rho_t=5.5$; the theoretical calculation does not impose 
a lower limit on the amplitude of signals in single detectors contributing coherently to the 
significance of an event.

We evaluate the accuracy of the recovered (observed) chirp mass $\mathcal{M}^z$ via the 
discrepancy $(\mathcal{M}_{\rm rec}^z - \mathcal{M}_{\rm inj}^z)/\mathcal{M}_{\rm inj}^z$ 
plotted in Figure~\ref{cbc_efficiency}, lower left plot.\footnote{The fractional error 
in observed chirp mass $\mathcal{M}^z$ is mathematically identical to the fractional error 
in physical chirp mass $\mathcal{M}$.} 
The vast majority of events found have a well recovered chirp mass with an accuracy better 
than 0.5\%, even for the small number of sources recovered at redshift $z>4$: 
the number of wrongly found injections with violently inaccurate 
$\mathcal{M}$ is order(10). The chirp mass is the chief parameter governing the 
frequency evolution of a compact binary system due to its emission of energy in GW, 
thus we can deduce the luminosity in GW of such systems with good accuracy 
out to extremely large distances. But note again that we cannot determine the system's
\emph{physical} masses without an independent determination of its redshift. 

The distribution of errors in observed total mass $M^z$ is significantly broader: see 
Figure~\ref{cbc_efficiency}, lower right plot, where there is a slight overall bias towards 
overestimating $M^z$ and a small population of injections for which the total mass 
is overestimated by 5--10\%. We find that this population consists of nearly equal-mass 
binaries which are found with somewhat more asymmetrical templates. 
Since the inspiral signal is significantly less sensitive to changes in mass ratio or total 
mass keeping a fixed $\mathcal{M}$ than vice versa, we expect a larger spread in 
recovered $M$ values than in $\mathcal{M}$. We might also expect a bias in the recovered
mass parameters due to taking the maximum SNR value over some region of the template bank, 
since the density of templates 
is greater at smaller $\mathcal{M}$ and at smaller $\eta$. 
The recovered mass ratio is also expected to be more sensitive to systematic differences in 
injected vs.\ template waveforms.

\subsection{Stochastic}
\begin{figure*}
\centering
\includegraphics[angle=000,width=0.44\textwidth,trim=0 -0.4cm 0 0.2cm,clip]{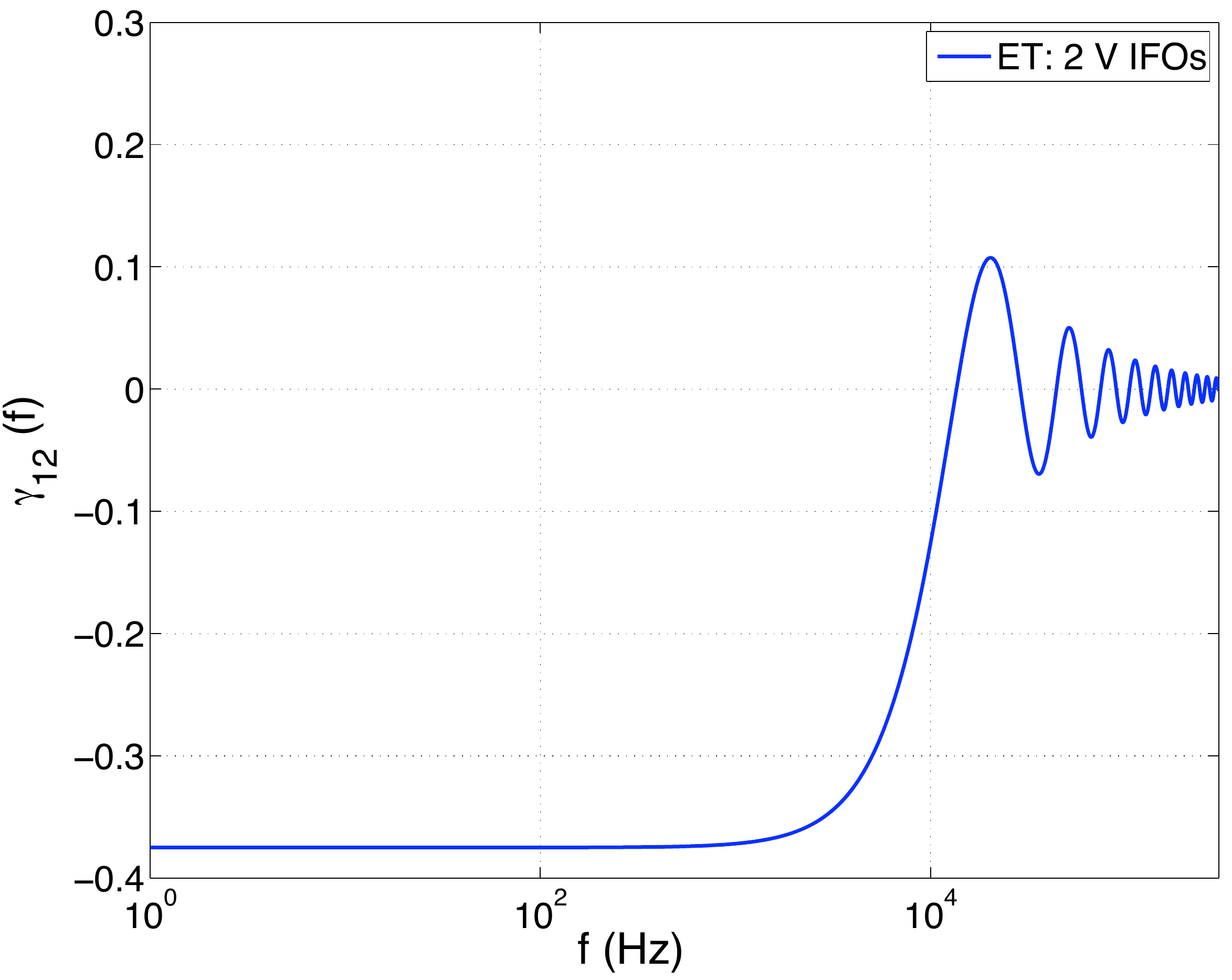}
\hspace*{0.4cm}
\includegraphics[angle=000,width=0.51\textwidth,trim=0 0.2cm 0 0,clip]{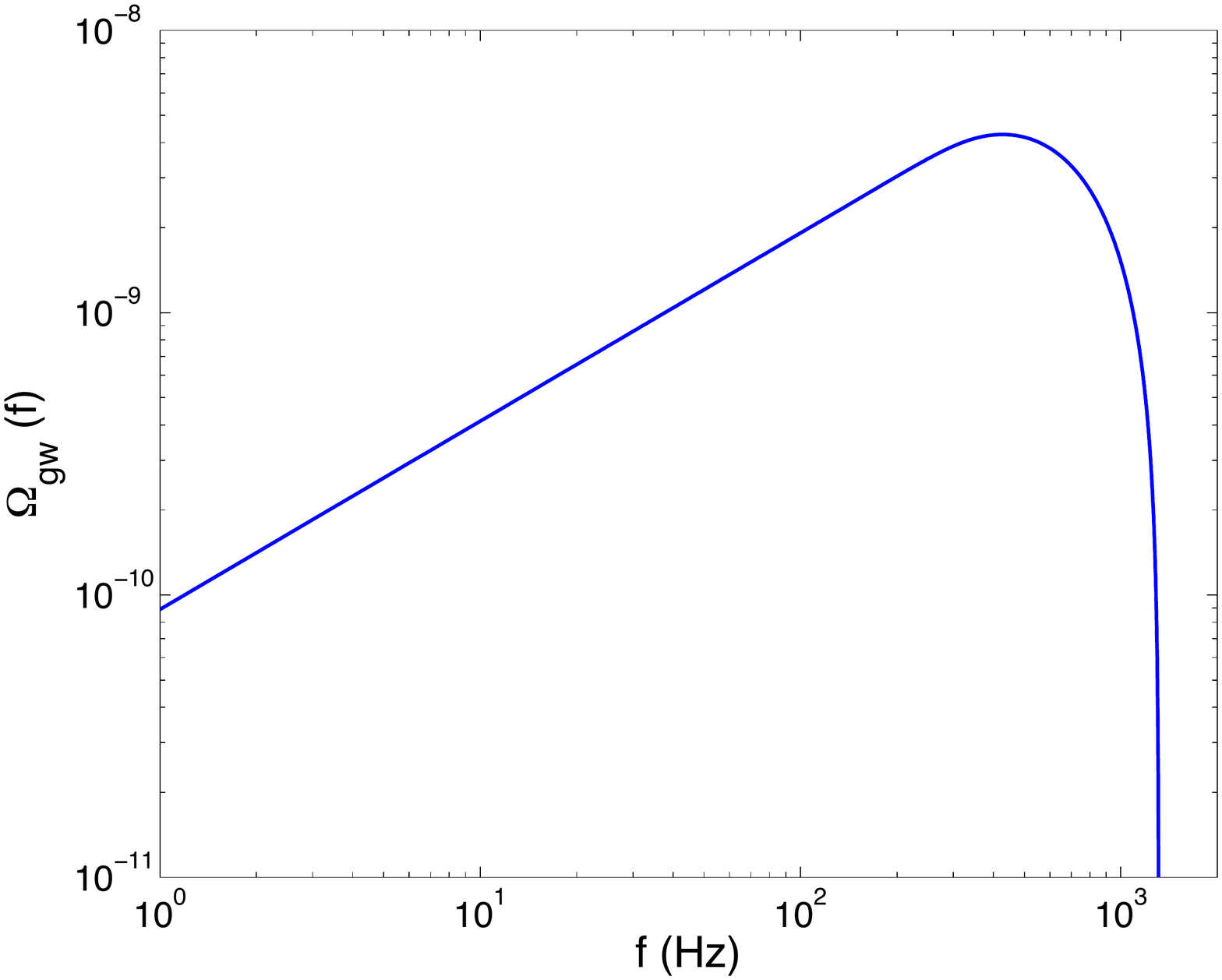}
\caption{
{\em Left}---Overlap reduction function for two V-shaped ET 
detectors separated by 120 degrees.  {\em Right}---Energy 
density parameter of the background produced by the coalescence 
of binary neutron stars, as a function of observed frequency.}
\label{fig-overlap}
\end{figure*}

The superposition of the GW signals from our population of BNS creates a background 
which is expected to be isotropic (the source position in the sky and polarization were 
selected from a uniform distribution) and stationary (the length of the data is much greater 
than the time interval between successive events, and the duration of the waveform).
Its properties in the frequency domain can be characterized by the dimensionless energy 
density parameter \cite{all99}:
\begin{equation}
\Omega_{gw}(f)=\frac{1}{\rho_{cr}}\frac{d\rho_{gw}}{d\ln f},
\end{equation}
where $\rho_{\rm gw}$ is the gravitational energy density and $\rho_{cr}=
3c^2H_0^2/(8\pi G)$ the critical energy density needed to make the Universe flat today.

This quantity is related to the one-sided ($f>0$) power spectral density in gravitational 
waves, at the detector output ($S_h$) :
\begin{equation}
\Omega_{gw}(f) = \frac{10 \pi^2}{3 \sin^2(\gamma) H_0^2} f^3 S_h(f)
\end{equation}
where $\gamma$ is the opening angle of the interferometer arms.

Note that the background from BNS is not Gaussian at frequencies $>10$ Hz, in the 
sense that the number of sources overlapping at a given time is too small for the central limit 
theorem to apply and for the distribution of the sum of the amplitudes to have a Gaussian 
distribution. Thus, knowledge of $\Omega_{gw}(f)$ does not \emph{completely} specify the 
statistical properties of the background, as there may be non-vanishing moments other 
than the variance. In particular, the amplitude distribution of the GW signal may exhibit large 
tails compared to the Gaussian case. 

For the population of neutron stars distributed according to the probability distributions 
discussed in Sec.III.A for the mass, redshift, position in the sky, polarization and inclination, 
the predicted $\Omega_{gw}$ is shown in Fig.~\ref{fig-overlap}, right panel, and can be 
derived from the expression \cite{reg11,zhu11,ros11}:
\begin{equation}
\Omega_{gw}(f)=\frac{1}{\rho_{cr} c} f F(f)
\label{eq:omega}
\end{equation}
where the integrated flux at the observed frequency $f$ is given by the sum of all the 
individual contributions at all redshifts:
\begin{equation}
F(f)=\int_0^{z_{\max}}  \frac{{\rm d}z}{4 \pi D_{\rm L}^2(z)} \frac{dE_{gw}}{df}(f,\bar{\mathcal{M}}(1+z))\frac{dR}{dz}(z)
\label{eq:flux}
\end{equation}
where $D_{\rm L}$ is the luminosity distance, $\frac{dE_{gw}}{df}$ the spectral energy density 
averaged over orientation and $\bar{\mathcal{M}}$ is the average physical chirp mass 
of the population. 

In the quadrupolar approximation, and assuming a circular orbit, 
\begin{equation}
 \frac{dE_{gw}}{df}(f,\mathcal{M}^z)= \frac{(G \pi)^{2/3} (\mathcal{M}^z)^{5/3}}{3} f^{-1/3},\ \mathrm{for}\,\ f<f_{lso}^z
\end{equation}
where $f_{lso}^z=(1+z)^{-1}f_{lso}$ is the observed (redshifted) frequency at the last stable orbit. 
The predicted energy density parameter increases as $f^{2/3}$ before it reaches a maximum 
$\Omega_{gw} \sim 4 \times 10^{-9}$ at around 600\,Hz, with a reference value at 100\,Hz of 
$\Omega_{\rm ref} = 1.9 \times 10^{-9}$.

The strategy to search for a Gaussian (or continuous) background, which could be confused 
with the intrinsic noise of a single interferometer, is to cross-correlate measurements of multiple 
detectors. When the background is assumed to be isotropic, unpolarized and stationary, the 
cross-correlation product is given by 
\cite{all99}
\begin{equation}
Y=\int_{0}^\infty \tilde{x}_1^*(f)\tilde{Q}(f)\tilde{x}_2(f)\, {\rm d}f
\label{eq:ccstat}
\end{equation}
and the expected variance, which is dominated by the noise, by
\begin{equation}
\sigma^2_Y \simeq \int_{0}^\infty S_n^1(f)S_n^2(f)|\tilde{Q}(f)|^2\,{\rm d}f,
\label{eq:ccvar}
\end{equation}
where
\begin{equation}
\tilde{Q}(f)\propto \frac{\gamma_{12} (f) \Omega_{\rm gw}(f)}{f^3S_n^1(f)S_n^2(f)}
\end{equation}
is a filter that maximizes the signal-to-noise ratio ($S/N$). In the above equation, $S_n^1$ 
and $S_n^2$ are the one-sided power spectral noise densities of the two detectors and 
$\gamma_{12}$ is the normalized overlap reduction function, characterizing the loss of 
sensitivity due to the separation and the relative orientation of the detectors: see 
Fig.~\ref{fig-overlap}, left panel. For two V-shaped detectors ($\gamma = \pi/3$) separated 
by $\beta = 2 \pi/3$ degrees, $\gamma_{12}(0) = \sin^2(\gamma) \cos(2 \beta) = -3/8$. The 
normalization ensures that $\gamma_{12} = 1$ for co-located and co-aligned L-shaped 
detectors. 

We analyzed the data with the cross-correlation code developed by the LIGO stochastic 
group. The data were split into $N=40320$ segments of length $T_{seg}=60$\,s, and for 
each segment the cross-correlation product and the theoretical variance were calculated 
using a template $\Omega \sim f^{2/3}$ in the range $10-500$\,Hz. The frequency 
resolution of our analysis was 0.25\,Hz. The final point estimate at 100\,Hz is given by 
\cite{S4stoch,S5stoch}
\begin{equation}
\hat{\Omega} = \frac{Y_{opt}}{T_{seg} \sum_i \sigma_{Y,i}^{-2}} 
\end{equation}
where $Y_i$ and $\sigma_{Y,i}^2$ are the cross-correlations and variances calculated for
each segment via Eq.~\eqref{eq:ccstat}, \eqref{eq:ccvar} respectively, and $Y_{opt}$ is 
the weighted sum
\begin{equation}
Y_{opt} = \sum_i Y_i\, \sigma_{Y,i}^{-2}.
\end{equation}
The standard error on this estimate is given by
\begin{equation}
\sigma_\Omega = \left[\sum_i \sigma_{Y,i}^{-2}\right]^{-1/2}  T_{seg}^{-1}.
\end{equation}
We found a point estimate at 100\,Hz of $2.00 \times 10^{-9}$ for the pair E1-E2, $1.90 \times 10^{-9}$ for E2-E3 
and $2.03 \times 10^{-9}$ for E2-E3 (an average of $\sim 1.97 \times 10^{-9}$), with error $\sigma_\Omega =  4.96 
\times 10^{-12}$ for the three pairs, at 100\,Hz, which corresponds to the analytical 
expectation of $\sim 1.9 \times 10^{-9}$ with a precision better than $5 \%$. 
Even if the 
background from compact binaries is not a \emph{Gaussian continuous} stochastic 
background, but rather a \emph{popcorn-like} background in the considered frequency 
range $f>10$\,Hz \cite{cow06,reg09,ros11}, our analysis has shown that non-Gaussian 
regimes can still be recovered by the standard cross-correlation statistics,
 confirming the results of \cite{dra03}.

\section{Future Development}
This first set of Mock Data included only a single type of signal, although the BNS 
systems we simulated are expected to be the most numerous and can thus yield 
much interesting information for astrophysics and cosmology.
Moreover
due to computational limitations we did not extend the simulations below a frequency
of 10\,Hz, though doing so might significantly improve the ability to extract signal
parameters. Future Mock Data sets should address these and other points by: 
\begin{enumerate}
 \item Including more types of GW sources;
 \item Using more complete or realistic waveforms;
 \item Using a more sophisticated noise model.
\end{enumerate}
Under the first heading, binary coalescence signals including stellar mass or 
intermediate mass black holes (IMBH) \cite{Huerta:2010un,Huerta:2010tp,
Gair:2010dx,AmaroSeoane:2009ui} are of particular interest. A small number of 
burst sources such as Type II supernovae are expected in the ET dataset and 
numerical simulations (for instance \cite{ott}) could be used to produce injection waveforms. 
It is also possible that primordial stochastic GW backgrounds exist in the ET sensitive
band \cite{DSD}; detecting these and determining their parameters would be an
interesting challenge given the significant contribution of astrophysical sources. 
For BNS coalescences, our injected waveforms could be improved by extending the
lower frequency cutoff, but also by modelling the merger phase (which depends 
strongly on the equation of state of NS matter, as well as the component masses). 
We expect that significant science can be extracted from BNS mergers, and from 
the tidal deformations occurring in the pre-merger phase, that are neglected in the
PN waveform model we currently use \cite{rmsucf09,Hinderer:2009ca}. 
Finally, we can simulate more realistic noise by adding occasional random glitches 
to the data, which may be supposed to be of instrumental or environmental origin.
To create an interesting challenge, single-detector glitches should be added 
with a higher rate than detectable signals. 

\subsection{Challenges for CBC analysis}
\label{sec:CBCfuture}
The initial search for coalescing binaries presented here, although moderately 
efficient below $z=1$, has some significant drawbacks. Here we discuss how it 
could be improved, and point to some current developments in CBC data analysis. 

In order to realize the full potential of ET's low frequency sensitivity down to 10\,Hz 
and below, waveforms lasting on the order of an hour or more should be matched 
filtered. For this a simple template bank as used in current searches would be 
prohibitively computationally costly \cite{Abbott:2005pf,por10} 
containing hundreds of thousands of templates or more. 
Currently, multi-band filter methods are being developed \cite{Cannon:2011vi} which 
split up the waveform into time slices with different frequency content: thus the earlier 
part of the waveform can be downsampled, reducing computational load. The resulting
template banks for each time slice are still large, and can be significantly reduced
by singular value decomposition \cite{Sathyaprakash:2003ua,Cannon:2011xk} 
allowing a computationally realistic search to be performed, while retaining the 
ability to reconstruct the SNR for each of the original templates.

We saw that the sensitivity of the coincident analysis was limited by the 
SNR threshold applied to single-detector triggers. 
Due to this threshold, signals from distant sources were often seen as double 
coincidences, which compete with a much larger noise background rate than triples. 
However, on lowering the threshold, the computational load would increase, as would
the number of background triple coincidences. As discussed earlier, recombining the 
three detector outputs into synthetic $+$ and $\times$ data streams should improve the 
overall separation of signal vs.\ noise, 

If detectors at other locations are operating at the same time as ET, a \emph{coherent} 
search should be performed to maximize sensitivity; such searches are currently under 
development, although facing an obstacle in their computational costs. 

We did not implement the null stream estimate of Eq.~(\ref{eq:nullpsd}) for the single-detector 
PSD within the CBC analysis. The difference with respect to the individual detector 
PSDs, including the contribution of signals, was less than 1\%, which we do not expect
to cause a measurable change in efficiency; however, if the contributions of GW signals 
were significantly higher, it might be beneficial to use the null stream PSD for template 
placement and matched filtering.

To obtain an unbiased estimate of the source parameters for each signal, a Bayesian 
analysis of the strain data should be performed 
\cite{Rover:2006bb,Rover:2007ij,Veitch:2008wd,Veitch:2009hd}. The chief conceptual 
challenge is the likely presence of \emph{many} signals within any stretch of data longer 
than about a minute \cite{reg09}. Na{\" \i}vely, in order to model them one would have to 
multiply the dimensionality of the source parameter space by the number of signals, 
however more efficient methods should exist; the problem is analogous to one faced in 
identifying multiple galactic binary sources in mock LISA data \cite{babak08} and 
\cite{por10} suggested that similar algorithms, for instance Markov Chain Monte Carlo 
based codes, could be used for ET. There will also be computational challenges 
in performing the analysis on hour-long stretches of data.

A conceptually difficult problem, not present in the current set of ET mock data, is 
to identify signals among an unmodelled background of non-Gaussian noise transients 
when the rates of signals and transients may both be large. As seen in the initial CBC 
analysis, the method of background estimation via time shifts between detectors is 
invalid if signals are frequent. The broad sensitivity spectrum and increased length 
of binary coalescence signals visible in ET gives us hope that signal-based vetoes 
based on the distribution of matched filter power over frequency \cite{Allen:2004gu} 
or over other parameters will be effective in separating signals from noise transients. 
The \emph{null stream} may also be useful to identify times when non-Gaussian noise is 
likely to produce loud false events, and to down-rank or veto such events.

If there are common non-Gaussian noise transients in more than one ET detector, which
may be caused by environmental disturbances, distinguishing these from GW signals 
may be more difficult, though it is still unlikely for such disturbances to cancel 
completely in the null stream. In any case, the use of signal-based vetoes should 
greatly assist in mitigating the effect of common noise for long-lived signals such as 
those from binary neutron stars.

Current methods for optimizing such vetoes involve adding simulated signals to strain 
data which are assumed \emph{not} to contain real signals; these must be revisited for 
ET, for example by using the null stream for simulations. Single-detector triggers
which fail a coincidence test could also be used to train glitch rejection methods.

One way to interpret such methods is to define a detection statistic for candidate 
events, with larger values indicating greater likelihood of signal \emph{vs.}\ noise, 
for instance the ``re-weighted SNR'' of \cite{Abbott:2007ai,s6-lowmass}. Under the 
weak assumption that some number of loud signals can be detected with high confidence, 
we should see an astrophysical event distribution over the statistic value of predictable 
form, superimposed on a population of noise transients. If the noise distribution is 
sufficiently different from that of signals, ideally decreasing rapidly at high statistic 
values \cite{s6-lowmass}, it may be possible to separate the two populations simply by 
fitting the astrophysical component. However, such a procedure would depend on the noise 
event population being sufficiently well understood.

\subsection{Challenges for stochastic background analysis}
According to various cosmological scenarios, we are bathed in a stochastic 
background of gravitational waves, memory of the first instant of the Universe, 
up to the limits of the Planck era and the Big Bang, and often seen as the Grail 
of GW astronomy. Proposed theoretical models include the amplification of 
vacuum fluctuations during inflation, pre-Big-Bang models, cosmic strings or 
phase transitions (see \cite{mag00,DSD}). In addition to the cosmological 
background (CGB), an astrophysical contribution (AGB) may have resulted 
from the superposition of a large number of unresolved sources since the 
beginning of stellar activity \cite{reg11}. In the range of terrestrial detectors (up 
to $f\sim 1$\,kHz) the AGB is expected to be dominated by the cosmological 
population of compact binaries, in particular BNS, and could be a noise that 
would mask the background of cosmological origin.

In this paper, we assume that the three ET detectors were independent and 
thus had no common (correlated) noise. A crucial prerequisite to searching ET
data for stochastic GW will be to identify and remove environmental noise 
that can corrupt the result of cross-correlation analysis with co-located 
detectors. Relevant methods are under development for the two co-aligned 
and co-located LIGO Hanford detectors \cite{fot08}.

One of the most important future tasks will be to subtract the astrophysical
contribution in order to allow detection of the primordial background. 
This could be done either in the frequency domain by modeling the power 
spectrum with high accuracy from theoretical studies, or characterizing its shape 
using Bayesian analysis of the data \cite{rob08}, or in the time domain by 
removing individual sources as previously discussed. 

The nature of the AGB may also differ from its cosmological counterpart, which is
expected  to be stationary, unpolarized, gaussian and isotropic, by analogy with 
the cosmic microwave background. On the one hand, the distribution of galaxies 
up to 100\,Mpc is not isotropic but strongly concentrated in the direction of the VIRGO 
cluster and the Great Attractor, and on the other hand, depending whether the time 
interval between events is short compared to the duration of a single event, the 
integrated signal may result in a continuous, popcorn noise or shot noise 
background~\cite{reg09}. 

In this paper we used the standard cross-correlation method for detection of stochastic
GW background, but new techniques exist or are under development in the 
LIGO/Virgo community to search for non-isotropic \cite{mit08,dip11} or non-gaussian 
stochastic backgrounds \cite{dra03,mar12}, and they will be tested in future challenges. 

Finally, the astrophysical background is not only a noise but it could carry crucial 
information about the star formation history, the metallicity, the mass range of neutron 
star and black hole progenitors, their physical properties, the rate of compact binary 
mergers: developing methods for parameter estimation will represent another 
important task in future challenges.

\section{Conclusion}

We have described the generation and first analyses of a mock data set for the 
proposed Einstein Telescope gravitational-wave observatory, containing a population 
of binary neutron star (BNS) inspiral signals at cosmological distances. Our motivation
for this MDC is both for data analysis, to consider the different challenges encountered
for data containing frequent and strong signals, and to emphasize \emph{science 
challenges} in relating the results of data analysis to outstanding questions in 
fundamental physics, astrophysics and cosmology~\cite{DSD}.

The challenge carried out in this paper is, in many ways, similar to the
Mock LISA Data Challenge \cite{rob08}, but there are some important technical differences.
In the case of LISA, the data analysis problem is not CPU or memory intensive.
Even year long signals at a frequency of $10^{-3}$\,Hz have only tens of thousands
of samples. In the case of ET, however, CPU and memory limit what problems
current algorithms are able to address. With the software and computer 
infrastructure that is presently available, it is impossible to address
the problem of ET data analysis to the fullest extent. For example, a
binary neutron star starting at 1\,Hz will last for about 5 days and there
is no way to filter such long signals with the matched filtering algorithms
accessible to us. It is necessary to explore and develop new search algorithms 
which don't require the entire template to be available at once to carry out
a search. More importantly, future MDCs focus on the challenge of
extracting useful science from ET, not just extraction of GW signals.

The design topology of the Einstein Telescope allows the construction of a unique null 
stream~\cite{Freise:2008dk} independent of the sky position. 
We have demonstrated that it is possible to 
recover the average spectrum of the GW signals by subtracting the ``pure noise'' power 
spectral density (PSD) obtained from the null stream, from the PSD in each individual 
detector. The recovered ``residual'' PSD has a power-law character extremely close to 
the $f^{-7/3}$ behaviour expected for inspiraling binary systems. The residual PSD can 
either be used as a diagnosis tool for future Mock Data Challenges and stochastic 
analyses, or as a research tool complementary to a more traditional stochastic analysis.  

The null stream is also expected to be a powerful tool for identifying and vetoing 
non-Gaussian features in the detector outputs; however, since the current set of ET mock 
data does not include such noise features, this use of the null stream will be a topic for 
future investigation. 

The analysis 
used to detect coalescing binary signals was similar to current pipelines employed in 
searching LIGO-Virgo data, and was able to recover a large fraction of simulated 
signals at redshifts approaching unity. Some signals were recovered up to redshifts
greater than 3 with good ($<1\%$) accuracy on chirp mass (the chief parameter 
determining the frequency evolution of inspiral signals). Overlap between two or more 
signals only rarely affected the performance of the analysis; however this could 
become a more critical issue if the lower frequency cutoff (taken to be 25\,Hz for the 
first CBC analysis) were reduced. 

We also searched for the GW background created by the superposition of all the binary 
inspiral signals up to a redshift of $z \sim 6$ using the standard cross-correlation 
statistic, considering the frequency range $10-500$\,Hz where the spectrum can be well 
approximated by a power law $\Omega_{gw}(f) \propto f^{2/3}$. 
Our point estimate at 100\,Hz is in good agreement with the analytical expectation 
(with a precision better than $\sim 5\%$), and our analysis shows that non-Gaussian 
regimes can be probed by the standard cross-correlation statistics near optimal 
sensitivity, confirming the work of \cite{dra03}.

Future mock data will include a wider range of signals, 
encompassing CBC signals from BNS, NSBH/BBH, IMBH systems; a possible primordial 
stochastic background; and rare burst-like signals such as core-collapse supernovae. 
The challenge will be not only to detect these signals but to measure their parameters, 
and ultimately to extract a unique range of information about astrophysics, fundamental 
physics and cosmology from the data.

Information on future challenges, and on how to participate will be posted on the ET 
MDC website \url{http://www.oca.eu/regimbau/ET-MDC_web/ET-MDC.html}. 

\section*{Acknowledgements}
We are grateful to Andreas Freise, Stefan Hild, Harald Lueck and particularly
Jolien Creighton, for a careful reading of the manuscript and useful comments. 
We thank the Albert Einstein Institute in Hannover, supported by the
Max-Planck-Gesellschaft, for use of the Atlas high-performance computing
cluster in the data generation and analysis, and Carsten Aulbert for
technical advice and assistance.
WDP, TGFL, and CVDB are supported by the research programme of the Foundation 
for Fundamental Research on Matter (FOM), which is partially supported by the 
Netherlands Organisation for Scientific Research (NWO). SG acknowledges support 
from NSF grant PHY-0970074 and UWM's Research Growth Initiative.
BSS, CR and TD were funded by the
Science and Technology Facilities Council (STFC) Grant No. ST/J000345/1 and
European Community's Seventh Framework Programme (FP7/2007-2013) under grant
agreement n 211743. 
CR was supported at Cardiff as a participant in an IREU program funded by NSF 
under the grant PHY-0649224 to the University of Florida.
KW's visit to Cardiff was supported by the International Work Experience for 
Technical Students, UK, programme for 2010.


\begin{thebibliography}{99}

\bibitem{GEO} B.~Willke \emph{et al.}, Class.\ Quant.\ Grav.\ {\bf 24}, S389 (2007).

\bibitem{LIGO} B.~Abbott \emph{et al.}, Rept.\ Prog.\ Phys.\ {\bf 72}, 076901 (2009). 

\bibitem{Virgo} F.~Acernese \emph{et al.}, AIP Conf.\ Proc.\ {\bf 794}, 307 (2005)

\bibitem{crab} B.~Abbott \emph{et al.}, Astrophys.\ J.\ {\bf 683}, L45 (2008).

\bibitem{stochNature} B.~Abbott \emph{et al.}, Nature {\bf 460}, 990 (2009).

\bibitem{ratespaper} J.~Abadie \emph{et al.}\ (LIGO Scientific Collaboration and 
Virgo Collaboration), Class.\ Quant.\ Grav.\ {\bf 27} 173001 (2010).

\bibitem{ET} M.~Punturo \emph{et al.}, Class.\ Quant.\ Grav. \ {\bf 27} 194002 (2010).

\bibitem{Abadie:2010yb}
J.~Abadie \emph{et al.}\ (LIGO Scientific Collaboration and Virgo Collaboration),
Phys.\ Rev.\ D {\bf 82} (2010)  102001 [arXiv:1005.4655].

\bibitem{Sathyaprakash:2011bh}
B.~Sathyaprakash, M.~Abernathy 
\emph{et al.}, ``Scientific Potential of Einstein Telescope,''
Proceedings of Rencontres de Moriond (2011), \emph{Gravitational Waves and Experimental Gravity}, March 21-27,
La Thuile, Italy [arXiv:1108.1423].

\bibitem{vdb10} C.~{Van Den Broeck}, to appear in the Proceedings of the 12th Marcel Grossman Meeting, 
Paris, 2009 [arXiv:1003.1386].

\bibitem{rmsucf09} 
  J.\,S.~Read, C.~Markakis, M.~Shibata, K.~Uryu, J.\,D.\,E.~Creighton and J.\,L.~Friedman,
  Phys.\ Rev.\  D {\bf 79} (2009) 124033
  [arXiv:0901.3258].

\bibitem{Hinderer:2009ca}
  T.~Hinderer, B.\,D.~Lackey, R.\,N.~Lang and J.\,S.~Read,
  Phys.\ Rev.\  D {\bf 81} (2010) 123016
  [arXiv:0911.3535].

\bibitem{Sathyaprakash:2009xt}
  B.\,S.~Sathyaprakash, B.\,F.~Schutz and C.~{Van Den Broeck},
  Class.\ Quant.\ Grav.\  {\bf 27} (2010) 215006 [arXiv:0906.4151].


\bibitem{zvbl10} W.~Zhao, C.~{Van Den Broeck}, D.~Baskaran, and T.G.F.~Li, Phys.\ Rev.\ D {\bf 83}, 023005 (2011).

\bibitem{Messenger:2011gi}
C.~Messenger and J.~Read, 
arXiv:1107.5725.

\bibitem{privateMessenger}
C.~Messenger, private communication (2012).

\bibitem{reg09} T.~Regimbau and S.\,A.~Hughes, Phys.\ Rev.\ D {\bf 79} 062002 (2009).

\bibitem{DSD}
M.~Abernathy \emph{et al.},
``Einstein gravitational wave Telescope: Conceptual Design Study'', 
European Gravitational Observatory document number ET-0106A-10, 
\url{http://www.et-gw.eu/etdsdocument}. 

\bibitem{Huerta:2010un}
E.\,A.~Huerta and J.\,R.~Gair,
Phys.\ Rev.\  D {\bf 83} (2011) 044020 [arXiv:1009.1985].

\bibitem{Huerta:2010tp}
E.\,A.~Huerta and J.\,R.~Gair,
Phys.\ Rev.\  D {\bf 83} (2011) 044021 [arXiv:1011.0421].

\bibitem{Gair:2010dx}
J.\,R.~Gair, I.~Mandel, M.\,C.~Miller and M.~Volonteri,
Gen.\ Rel.\ Grav.\  {\bf 43} (2011) 485 [arXiv:0907.5450].

\bibitem{AmaroSeoane:2009ui}
P.~Amaro-Seoane and L.~Santamaria,
Astrophys.\ J.\  {\bf 722} (2010) 1197 [arXiv:0910.0254].

\bibitem{ET-B}  S.~Hild \emph{et al.}, arXiv:0810.0604v2 (2008).

\bibitem{ET-C}  S.~Hild \emph{et al.}, Class.\ Quant.\ Grav.\ \textbf{27} (2010) 015003.

\bibitem{ET-D}  S.~Hild \emph{et al.}, Class.\ Quant.\ Grav.\ \textbf{28} (2011) 094013.

\bibitem{Hild:2011ub}
S.~Hild,
``Beyond the Second Generation of Laser-Interferometric Gravitational Wave Observatories,''
arXiv:1111.6277 [gr-qc].

\bibitem{aLIGO}
G.~M. Harry and the LIGO Scientific Collaboration, Class.\ Quant.\ Grav.\ \textbf{27} (2010) 084006;
Advanced LIGO Reference Design, LIGO Document
M060056-v2, \url{https://dcc.ligo.org/cgi-bin/DocDB/ShowDocument?docid=m060056}.
Advanced LIGO project URL is \url{https://www.advancedligo.mit.edu/.}

\bibitem{aVirgo}
G.~Losurdo and the Advanced Virgo Team, Virgo document VIR-0042A-07 (2007), \url{https://tds.ego-gw.it/ql/?c=1900} 

\bibitem{LIGODesign}
A. Abramovici \emph{et al.}, Science {\bf 256} 325 (1992).

\bibitem{VirgoDesign}
B. Caron \emph{et al.}, Class.\ Quant.\ Grav.\ {\bf 14} 1461 (1997).

\bibitem{Schutz:2011tw}
B.\,F.~Schutz,
Class.\ Quant.\ Grav.\  {\bf 28} (2011) 125023 [arXiv:1102.5421].

\bibitem{Abbott:2009tt}
  B.~P.~Abbott \emph{et al.}\ (LIGO Scientific Collaboration),
  Phys.\ Rev.\  D {\bf 79} (2009) 122001
  [arXiv:0901.0302].

\bibitem{Finn:1992xs}
L.~S.~Finn and D.~F.~Chernoff,
Phys.\ Rev.\ D {\bf 47} (1993) 2198
[gr-qc/9301003].

\bibitem{den} ET-B is one of the design sensitivity curves for Einstein Telescope
\cite{ET-B}. An analytical fit for ET-B can be found at:
\url{https://workarea.et-gw.eu/et/WG4-Astrophysics/base-sensitivity/}.

\bibitem{kal04} V.~Kalogera \emph{et al.}, Astrophys.\ J.\ {\bf 614} 137 (2004).

\bibitem{hop06} A.\,M.~Hopkins and J.~Beacom, Astrophys.\ J.\ {\bf 651} 142 (2006).

\bibitem{popsynth} T.\ Piran, Astrophys.\ J.\ {\bf 389}, L83 (1992); 
 A.\ V.\ Tutukov and L.\ R.\ Yungelson, Mon.\ Not.\ R.\ Astron.\ Soc.\ {\bf 268}, 871 (1994); 
 V.\ M.\ Lipunov \emph{et al.}, 
 Astrophys.\ J. {\bf 454}, 593 (1995); 
 S.\ Ando, J.\ Cosmology and Astroparticle Phys.\ {\bf 06}, 007 (2004);
 J.\ A.\ de Freitas Pacheco, T.\ Regimbau, A.\ Spallici, and S.\ Vincent, Int.\ J.\ Mod.\ Phys.\ D {\bf 15}, 235 (2006); 
 K.\ Belczynski \emph{et al.}, 
 Astrophys.\ J.\ {\bf 648}, 1110 (2006); 
 R.\ O'Shaughnessy, K.\ Belczynski, and V.\ Kalogera, Astrophys.\ J.\ {\bf 675}, 566 (2008).

\bibitem{bel} K.\ Belczynski and V.\ Kalogera, Astrophys.\ J.\ Lett.\ {\bf 550}, L183 (2001); 
 K.\ Belczynski \emph{et al.}, 
 Astrophys.\ J.\ {\bf 648}, 1110 (2006).

\bibitem{ber06} E.\ Berger et al., Astrophys.\ J.\ {\bf 664}, 1000
(2006).




\bibitem{Allen:2005fk}
B.~Allen, W.\,G.~Anderson, P.\,R.~Brady, D.\,A.~Brown and J.\,D.\,E.~Creighton,
arXiv:gr-qc/0509116.

\bibitem{Brown:2005zs}
  D.\,A.~Brown for the LIGO Scientific Collaboration,
  Class.\ Quant.\ Grav.\  {\bf 22} (2005) S1097
  [arXiv:gr-qc/0505102].

\bibitem{Pai:2000zt}
A.~Pai, S.~Dhurandhar and S.~Bose,
Phys.\ Rev.\  {\bf D64} (2001)  042004.

\bibitem{Pai:2001cf}
A.~Pai, S.~Bose and S.~Dhurandhar,
Class.\ Quant.\ Grav.\ {\bf 19} (2002) 1477.

\bibitem{Bose:2002by}
S.~Bose,
Class.\ Quant.\ Grav.\ {\bf 19} (2002) 1437.

\bibitem{Harry:2010fr}
  I.~W.~Harry and S.~Fairhurst,
  Phys.\ Rev.\ D {\bf 83} (2011) 084002
  [arXiv:1012.4939 [gr-qc]].
  
\bibitem{Bose:2011km}
S.~Bose, T.~Dayanga, S.~Ghosh and D.~Talukder,
Class.\ Quant.\ Grav.\ {\bf 28} (2011) 134009 [arXiv:1104.2650].

\bibitem{all99} 
  B.~Allen and J.\,D.~Romano,
  Phys.\ Rev.\ D {\bf 59} (1999) 102001.

\bibitem{mag00} 
  M.~Maggiore,
  Phys.\ Rept.\ {\bf 331} (2000) 283.

\bibitem{Buonanno:2002fy}
A.~Buonanno, Y.\,-b.~Chen and M.~Vallisneri,
Phys.\ Rev.\ D {\bf 67} (2003) 104025 [gr-qc/0211087].

\bibitem{approximants} A.~Buonanno, B.~Iyer, E.~Ochsner, Y.~Pan, and 
B.\,S.~Sathyaprakash, Phys.\ Rev.\ D {\bf 80} 084042 (2009).

\bibitem{Salgado:1994}
M. Salgado, S. Bonazzola, E. Gourgoulhon, and P. Haensel, Astronomy and Astrophysics, {\bf 291}, 
155 (1994).

\bibitem{Frasca:1997}
  S.~Frasca and M.A. Papa, {\em An untility for VIRGO data simulation, i.e.
  how to build noise data from the knowledge of the spectrum,}
  VIR-NOT-ROM-1390-090, Issue 1 (1997).

\bibitem{Freise:2008dk}
  A.~Freise, S.~Chelkowski, S.~Hild, W.~Del Pozzo, A.~Perreca and A.~Vecchio,
  Class.\ Quant.\ Grav.\ {\bf 26} (2009) 085012
  [arXiv:0804.1036].

\bibitem{trigscan_poster}
  C.~Robinson, A.~Sengupta and B.\,S.~Sathyaprakash, 
  \url{https://dcc.ligo.org/cgi-bin/DocDB/ShowDocument?docid=36649}.

\bibitem{ethinca}
  C.\,A.\,K.~Robinson, B.\,S.~Sathyaprakash and A.\,S.~Sengupta,
  Phys.\ Rev.\ D {\bf 78} (2008) 062002
  [arXiv:0804.4816].

\bibitem{reg11} T.~Regimbau, Res.\ Astron.\ Astrophys.\ {\bf 11} (2011) 369 [arXiv:1101.2762v3].

\bibitem{zhu11} 
  X.\,J.~Zhu, E.~Howell, T.~Regimbau, D.~Blair and Z.\,H.~Zhu,
  Astrophys.\ J.\ {\bf 739} (2011) 86
  [arXiv:1104.3565].

\bibitem{ros11}  
P.\,A.~Rosado,
Phys.\ Rev.\ D {\bf 84} (2011) 084004
[arXiv:1106.5795].

\bibitem{S4stoch}
B.~Abbott \emph{et al.} (LIGO Scientific Collaboration),
Phys.\ Rev.\ D {\bf 95} (2005) 221101.

\bibitem{S5stoch}
B.~Abbott \emph{et al.} (LIGO Scientific Collaboration),
Nature {\bf 460} (2009) 7258.

\bibitem{cow06} 
D.~Coward and T.~Regimbau, New Astronomy Reviews {\bf 50} (2006) 461.

\bibitem{ott} 
 C.\,D.~Ott, A.~Burrows, L.~Dessart and E.~Livne, Phys.~Rev.~Lett.\ {\bf 96} 
 (2006) 201102; 
 C.\,D.~Ott  \emph{et al.}, Class.\ Quant.\ Grav.\ {\bf 24} (2007) 139, and 
 Phys.~Rev.~Lett.\ {\bf 98} (2007) 261101; 
 K.~Kotake, W.~Iwakami, N.~Ohnishi and S.~Yamada, Astrophys.\ J.\ Lett.\ {\bf 697} (2009) 133; 
 A.~Marek, H.\,-T.~Janka and E.~Muller, Astron.\ Astrophys.\ {\bf 496} (2009) 475.

\bibitem{Abbott:2005pf}
B.~Abbott \emph{et al.}\ (LIGO Scientific Collaboration),
Phys.\ Rev.\  {\bf D72} (2005) 082002 [gr-qc/0505042].

\bibitem{por10} L.\ Bosi and E.\ K.\ Porter, arXiv:0910:0380.

\bibitem{Cannon:2011vi}
  K.~Cannon {\it et al.},
  arXiv:1107.2665 [astro-ph.IM].

\bibitem{Schutz:1986gp}
B.~F.~Schutz,
Nature {\bf 323} (1986) 310.

\bibitem{Sathyaprakash:2003ua}
B.\,S.~Sathyaprakash and B.\,F.~Schutz,
Class.\ Quant.\ Grav.\ {\bf 20} (2003) S209 [gr-qc/0301049].

\bibitem{Cannon:2011xk}
K.~Cannon, C.~Hanna and D.~Keppel,
arXiv:1101.4939.

\bibitem{Rover:2006bb}
C.~R{\" o}ver, R.~Meyer and N.~Christensen,
Phys.\ Rev.\ D {\bf 75} (2007) 062004 [gr-qc/0609131].

\bibitem{Rover:2007ij}
C.~R{\" o}ver, R.~Meyer, G.\,M.~Guidi, A.~Vicere and N.~Christensen,
Class.\ Quant.\ Grav.\ {\bf 24} (2007) S607 [arXiv:0707.3962].

\bibitem{Veitch:2008wd}
J.~Veitch and A.~Vecchio,
Class.\ Quant.\ Grav.\ {\bf 25} (2008) 184010 [arXiv:0807.4483].

\bibitem{Veitch:2009hd}
J.~Veitch and A.~Vecchio,
Phys.\ Rev.\ D {\bf 81} (2010) 062003 [arXiv:0911.3820].

\bibitem{babak08} 
S.~Babak \emph{et al.}, Class.\ Quant.\ Grav.\ {\bf 25} (2008) 184026.

\bibitem{Allen:2004gu}
  B.~Allen, 
  Phys.\ Rev.\ D {\bf 71} (2005) 062001.

\bibitem{Abbott:2007ai}
B.~Abbott \emph{et al.} (LIGO Scientific Collaboration),
Phys.\ Rev.\ D {\bf 78} (2008) 042002 [arXiv:0712.2050].

\bibitem{s6-lowmass}
  J.~Abadie \emph{et al.} (LIGO Scientific Collaboration and Virgo Collaboration), 
  Phys.\ Rev.\ D {\bf 85} (2012) 082002
  [arXiv:1111.7314].  

\bibitem{rob08} 
  E.\,L.~Robinson, J.\,D.~Romano and A.~Vecchio,
  Class.\ Quant.\ Grav.\ {\bf 25} (2008) 184019
  [arXiv:0804.4144].

\bibitem{fot08} 
N.~Fotopoulos, Journal of Physics: Conference Series {\bf 122} (2008) 012032. 

\bibitem{hog01} 
C.\,J.~Hogan and P.\,L.~Bender,  Phys.\ Rev.\ D (2001) {\bf 64} 062002. 

\bibitem{mit08}
  S.~Mitra \emph{et al.}, 
  Phys.\ Rev.\ D {\bf 77} (2008) 042002
  [arXiv:0708.2728].

\bibitem{dip11}
  D.~Talukder, S.~Mitra and S.~Bose,
  Phys.\ Rev.\ D {\bf 83} (2011) 063002.
  [arXiv:1012.4530].

\bibitem{dra03}
S.~Drasco and E.\,E.~Flanagan, Phys.\ Rev.\ D {\bf 67} (2003) 082003.

\bibitem{mar12}
L.~Martellini and T.~Regimbau, in preparation.


\end{thebibliography}
\end{document}